\documentclass[dvips,generic,preprint,12pt]{imsart}
\setattribute{journal}{name}{}

\renewcommand\baselinestretch{1.}

\RequirePackage[OT1]{fontenc}
\RequirePackage{amsthm,amsmath,natbib}
\RequirePackage[colorlinks,citecolor=blue,urlcolor=blue]{hyperref}
\RequirePackage{hypernat}
\RequirePackage{comment}
\RequirePackage{graphicx,subfigure,latexsym,amssymb}
\RequirePackage{float,epsfig,multirow,rotating,times}
\RequirePackage{upgreek,wrapfig}

\arxiv{math.PR/0000000}
\startlocaldefs
\newcommand {\ctn}{\citet} 
\newcommand {\ctp}{\citep}       

\numberwithin{equation}{section}
\theoremstyle{plain}

\newcommand{\bbeta}{\boldsymbol{\beta}}

\newcommand{\bxi}{\boldsymbol{\xi}}

\newcommand{\bSigma}{\boldsymbol{\Sigma}}
\newcommand{\bsigma}{\boldsymbol{\sigma}}

\newcommand{\bmu}{\boldsymbol{\mu}}

\newcommand{\bvarphi}{\boldsymbol{\varphi}}

\newcommand{\bOmega}{\boldsymbol{\Omega}}

\newcommand{\bB}{\boldsymbol{B}}
\newcommand{\bC}{\boldsymbol{C}}

\newcommand{\bh}{\boldsymbol{h}}
\newcommand{\bH}{\boldsymbol{H}}

\newcommand{\bM}{\boldsymbol{M}}
\newcommand{\bI}{\boldsymbol{I}}

\newcommand{\bA}{\boldsymbol{A}}
\newcommand{\bS}{\boldsymbol{S}}

\newcommand{\bQ}{\boldsymbol{Q}}

\newcommand{\bv}{\boldsymbol{v}}
\newcommand{\bs}{\boldsymbol{s}}

\newcommand{\bzero}{\boldsymbol{0}}

\newcommand {\mU}{\mathcal{U}}
\endlocaldefs

\begin{document}
\renewcommand\baselinestretch{1.}

\begin{frontmatter}
\title{Bayesian Nonparametric Estimation of Milky Way Parameters Using Matrix-Variate Data, in a New Gaussian Process Based Method}
\runtitle{Estimation using New ${\cal GP}$ Based Method}

\begin{aug}
\author{
{\fnms{Dalia} \snm{Chakrabarty}\thanksref{t1,m1}\ead[label=e1]{d.chakrabarty@warwick.ac.uk},\ead[label=e2]{dc252@le.ac.uk}},
{\fnms{Munmun} \snm{Biswas}\thanksref{t3,m3}\ead[label=e3]{munmun.biswas08@gmail.com}},
{\fnms{Sourabh} \snm{Bhattacharya}\thanksref{t4,m3}\ead[label=e4]{sourabh@isical.ac.in}}
},
\thankstext{t1}{Associate Research fellow at Department of Statistics,
  University of Warwick and Lecturer of Statistics at Department of Mathematics,
  University of Leicester}
\thankstext{t3}{PhD student in Statistics and Mathematics Unit, Indian Statistical Institute} 
\thankstext{t4}{Assistant Professor in Bayesian and Interdisciplinary Research Unit, Indian Statistical Institute}

\runauthor{Chakrabarty, Biswas and Bhattacharya}

\affiliation{University of Warwick \and University of Leicester, Indian statistical Institute}

\address{\thanksmark{m1} Department of Statistics\\
University of Warwick\\
Coventry CV4 7AL,
U.K.\\
\printead*{e1}\\
\and\\
Department of Mathematics\\
University of Leicester \\
Leicester LE1 7RH,
U.K.\\
\printead*{e2}
}

\address{\thanksmark{m3}
Indian Statistical Institute\\
203, B. T. Road\\ 
Kolkata 700108, India\\
\printead*{e3}
\printead*{e4}
}

\end{aug}

\begin{abstract} { 
  In this paper we develop an inverse Bayesian approach to find the
  value of the unknown model parameter vector that supports the real
  (or test) data, where the data comprises measurements of a
  matrix-variate variable. The
  method is illustrated via the estimation of the unknown Milky Way
  feature parameter vector, using available test and simulated
  (training) stellar velocity data matrices. The data is represented
  as an unknown function of the model parameters, where this
  high-dimensional function is modelled using a high-dimensional
  Gaussian Process (${\cal GP}$).  The model for this function is
  trained using available training data and inverted by Bayesian
  means, to estimate the sought value of the model parameter vector at
  which the test data is realised. We achieve a closed-form expression
  for the posterior of the unknown parameter vector and the parameters
  of the invoked ${\cal GP}$, given test and training data. 
  We perform
  model fitting by comparing the observed data with predictions made
  at different summaries of the posterior probability of the model
  parameter vector. As a supplement, we undertake a leave-one-out
  cross validation of our method. }
\end{abstract}



\end{frontmatter}

\renewcommand\baselinestretch{1.}
{
\section{Introduction}
\label{sec:intro}
Curiosity about the nature of the parameter space of the Milky Way
that we earthlings live in, is only natural. In this paper, we discuss
the learning of the parameters characterising those Milky Way features
that bear influence upon the motion of individual stars that lie in
the neighbourhood of the Sun. Astrophysical modelling indicates that
in the solar neighbourhood, effects of different features of the Milky
Way are relevant \ctp{minchev_quillen,chakrabarty07,barbara}. Such
features include an elongated bar-like structure made of stars (the
stellar bar) that rotates, pivoted at the centre of the Galaxy. In
addition, the spiral arms of the Galaxy are also relevant. Thus, the
motions of stars in the solar neighbourhood are affected by the
parameters that define these Galactic features. Included in these
feature parameters are the locations of the observer of such
motions--we from Earth observe such motions, so that the stellar
velocities are recorded to attain the observed values, given where in
the Galaxy we are measuring these velocities from. On astronomical
scales, the Earth's location in the Milky Way is equivalent to the
location of the Sun inside the Galaxy. Our location in the
two-dimensional (by assumption) Galactic disk, is given by the angular
separation of the Sun from a chosen line (an identified axis of the
aforementioned stellar bar) and the distance from the Sun to the
centre of the Galaxy. These two location parameters are the components
of the two-dimensional location $\bS$ of the observer. As motivated
above, parameters of the bar, spiral pattern and other Milky Way
features, can also affect the motions of stars that are observed. (See
section {\bf{S-1}} of the supplementary material for details). Given
that these galactic feature parameters affect the solar neighbourhood,
if motions of a sample of stars in this neighbourhood are measured,
such data will harbour information about these feature
parameters. Then, the inversion of such measured motions will in
principle, allow for the learning of the unknown feature
parameters. This approach has been adopted in the modelling of our
galaxy, to result in the estimation of the angular separation of the
Sun from a chosen axis of the bar, and the distance of the Sun from the
Galactic centre \ctp{minchev_bar,fux,dehnen,tremainewu}. The other
relevant feature parameters are typically held constant in such
modelling.

The above inverse problem is then an example application of the method of
science that is typified by attempts at learning the unknown model
parameter vector given observed data, where the causal relationship
between the observable and the model parameter vector $\bS$, is not
necessarily known. This unknown relationship or function, can itself
be learnt using available ``training data''. Once this function is
learnt, it can in principle be inverted to predict the
unknown value of $\bS$ at which the measured data--i.e. ``test
data''--is realised. Such test data is contrasted with ``training
data'', which is data generated at known or chosen values of
$\bS$ (for example, via simulations or obtained as archival data).

The learning of a high-dimensional function from available training
data, using standard nonparametric methods (such as spline fitting or
wavelet based learning) is expected to be unsatisfactory since
modelling high-dimensional functions using splines/wavelets may fail
to adequately take into account the correlation structure between the
component functions. Also, the complexity of the computational task of
learning the unknown function from the data--and in particular of
inverting it--only increases with dimensionality.  Furthermore, the
additional worry in the classical approach is that parameter
uncertainty is ignored, though the same can be addressed in a Bayesian
framework. An added advantage of the Bayesian approach is that priors
on the unknown parameters can bring in extra information into the
model, allowing for a training data set of comparatively
smaller size (than that required in the classical approach), to be
adequate.

Solving for the value of $\bS$ that supports the real or test data
requires operating the inverse of the learnt function on the test
data. The existence and uniqueness of such solution can be questioned
given that the problem may not even be well-posed in a Hadamard sense
\ctp{jlofillposed,hofmann,tarantola2005}. The problem may even be
ill-conditioned since errors in the measurement may exist. Such
worries about ill-posedness and ill-conditioning are mitigated in the
Bayesian framework \ctp{bayesinv,andrew}. In this approach, the 
solution entails computation of the posterior
probability of the unknown $\bS$ (at which the test data is
realised), given all data. Given the inherent
inadequacies of learning using splines/wavelets discussed above, we
opt to model the unknown functional relationship between data and
model parameter $\bS$ with a high-dimensional ${\cal GP}$.
Similarly, in our application of interest, the unknown functional relation
between the high-dimensional observations on stellar motions and the
unknown observer location vector $\bS$ is modelled as a
high-dimensional ${\cal GP}$.  In this exercise, Galactic feature
parameters other than the observer location are maintained as constants.

\ctn{chakrabarty07} constructed four different base-astronomical
models of the solar neighbourhood, each at a chosen value of the ratio
of the rate of rotation of the spiral pattern ($\Omega_s$) to that of
the bar ($\Omega_b$). Non-linear dynamical evolution of each of these
four base-astronomical models were carried out by \ctn{chakrabarty07},
resulting in four independent data sets, each consisting of $n$ blocks
of $j$ number of $k$-dimensional stellar velocity vectors, where each
block is generated at a chosen value of $\bS$ (aka, a ``design
point''). At each possible chosen location $\bs$ of the Sun, the
dynamical evolution of a given base-astronomical model of the Galaxy
generates a block representing the $k$-dimensional velocity of each of
$j$ stars, where these stars are chosen as neighbours of the
Sun. Thus, there are $n$ design points and each training data set
consists of $n$ number of $j\times k$-matrices, with a matrix
generated at the corresponding design point. There are four such
training data sets generated, by performing the evolution of each
base-astronomical model. In addition, there is a measured, stellar
velocity data matrix--of dimensionality $j\times k$ again--available,
but this time, we do not know what is the value of $\bS$ at which this
measured/test data has been realised. It is this unknown value of
$\bS$ that we seek to Bayesianly learn, given the test data and one
training data set at a time.

It maybe asked that if a stellar velocity matrix can be generated at a
chosen $\bs$, via the evolution of a base-astronomical model, does
this not amount to stating that the causal relationship between the
observable (velocity matrix) and model parameter ($\bS$) is already
known?  Indeed this knowledge must be embedded within the evolutionary
scheme implemented on any base-astronomical model. Thus, the forward
evolution of a base-astronomical model is possible (via Newton's
equations of motion), in order to generate a velocity matrix at a
chosen $\bs$. However the inversion of this evolution--aimed at
recovering the sought $\bs$ at which the measured velocity matrix is
generated--is not possible in general, owing to non-linear dynamical
effects, or chaos, that impede reversibility in evolution; see
\ctn{sengupta03}, Section~6.6 of \ctn{chakrabarty07}, Section~7 of
\ctn{fux}. The strength of such chaos is different in the different
base-astronomical models, caused by the different values of
$\Omega_s/\Omega_b$, (discussed below in
Section~\ref{sec:casestudy}). This difficulty of inversion triggers
the need to learn the inverse of the function that expresses the
observable as a function of $\bS$, independently from each of the four
available training data sets. This learnt inverse function is then to
be operated upon the measured (test) data to predict the value of
$\bS$ in the Milky Way, in each of the four cases that represent four
possible astronomical models of the Milky Way. We of course, predict
this value of $\bS$ Bayesianly, by using a high-dimensional
${\cal GP}$ to model the velocity data. We then achieve a closed-form
posterior probability density of the sought $\bs$ and relevant
parameters of this ${\cal GP}$, given the test and training
data. Marginal posterior distribution of the components of the sought
$\bs$ vector are inferred using MCMC, for each base-astronomical model
(i.e. each training data set) used. Our focus in this work is to make
inference on all values of $\bS$ at which the test data is realised,
in each of the four astronomical models of the Galaxy--selection of
the base-astronomical model is beyond the scope of this paper (see
Section~\ref{sec:results}).

In the astronomical literature, Milky Way feature parameters in the
solar neighbourhood have been explored via simulation based studies
\ctp{hydro,fux97} while similar estimation is performed using other
(astronomical) model-based studies \ctp{aumer,
  perrymanbook,golubovthesis}. \ctn{chakrabarty07} attempted
estimation of the sought Galactic parameters via a test of hypothesis
exercise: a non-parametric frequentist test was designed to test for
the null that the observed stellar velocity data matrix is sampled from the
estimated density of a synthetic velocity data matrix generated at the
corresponding chosen value of the Milky Way feature parameter vector
$\bS$. The $p$-value of the used test statistic was recorded for each
choice of $\bs$. The choices of $\bs$ at which the highest $p$-values
were obtained, were considered better supported by the observed
data. Hence the empirical distribution of these $p$-values in the
space of $\bS$, was used to provide interval estimates of the Milky
Way feature vector. However, this method required computational effort
and is highly data intensive since the best match is sought over a
very large collection of training data points. This shortcoming had
compelled \ctn{chakrabarty07} to resort to an unsatisfactory coarse
gridding of the space of ${\bS}$. This problem gets acute enough for
the method to be rendered useless when the dimensionality of the
vector ${\bS}$ that we hope to learn, increases. Moreover, the method
of quantification of uncertainty of the estimate of the location is
also unsatisfactory, dependent crucially on the binning details, which
in turn is bounded by cost and memory considerations.


In the method we develop here,
we demonstrate the effectiveness
of our Gaussian Process based method with much smaller data sets than
were used in the past. The other major advantage of this presented
method is that it readily allows for the expansion of dimensionality
of the model parameter vector and is capable of taking measurement
errors into account.

The rest of the paper is structured as follows. 
In Section \ref{sec:modelling}, we present the details of the
modelling strategy that we adopt.  The treatment of measurement errors
within the modelling is discussed in Section~\ref{subsec:measerr}.  In
Section~\ref{sec:casestudy} we discuss the application via which the
new method is illustrated while details of the inference are discussed
in Section~\ref{subsec:implementtmcmc}. Section~\ref{sec:results}
contains results obtained from using available real and training
data. We compare the obtained results with the estimates available in
the astronomical literature in
Section~\ref{sec:compare}. Section~\ref{sec:modelfit} presents results of
model fitting by comparing test data with predictions made at
different summaries of the posterior of the model parameter vector
$\bS$.  The paper is rounded up with Section~\ref{sec:discussions}.

}
\section{Model}
\label{sec:modelling}
\noindent
In this section we discuss the generic methodology that we use to
learn the unknown location vector of the observer in the Milky Way
disk, given the matrix-variate test and training stellar velocity
data. Once the method is motivated, we implement it in the following
section, to perform the learning relevant to the application at hand.

If a matrix-variate observable is expressed as an unknown
matrix-variate function of the model parameter $\bS$, and this unknown
causal relationship between observable and $\bS$ is modelled by a
matrix-variate Gaussian Process (${\cal GP}$), it would imply that one
realisation from such a matrix-variate ${\cal GP}$ would be a set of
the observed matrices that will be jointly distributed as 3-tensor
normal, parametrised by a mean matrix and 3 covariance matrices
\ctp{hoff2011}. While applications of the same are being developed
(Wang $\&$ Chakrabarty), here we undertake an alternative and
equivalent modelling strategy. We vectorise our intrinsically
matrix-variate data sets to achieve a close-form expression for the
joint posterior probability of the unknown parameters that we are
interested in learning from the data. This leads to the functional
relationship between the data and model parameter vector being
rendered vector-valued, modelled by a vector-variate ${\cal GP}$, a
set of realisations from which is jointly matrix normal, parametrised
by matrix-variate parameters that we intend to learn from the data,
along with the unknown $\bs$ at which the measured data is realised.

Let $j$ number of measurements of a $k$-dimensional variable be
available; this vector variable is referred to below as the
``observable''. Thus the measurements of this observable constitute a
$j\times k$-dimensional matrix. We refer to the measured data as test
data and seek the unknown value $\bs^{(new)}$ of model parameter $\bS$
at which it is realised. Let data be generated at $n$ known values of
${\bS}$: ${\bs}_1^\star,\ldots,{\bs}_n^\star$. Then
$\{{\bs}_1^\star,\ldots,{\bs}_n^\star\}$ is the design set and
$\bs_i^\star$ is the $i$-th {design vector} at which the $i$-th
{synthetic data matrix} is generated, $i=1,2,\ldots,n$.  Then
these $n$ synthetic data matrices comprise a {training data} set. Here a
data matrix is $j\times k$-dimensional. As motivated in the
introductory section, we express the relation between the observable ${\bf
  V}$ and unknown model parameter vector $\bS$ as ${\bf
  V}={\bxi}(\bS)$, where $\bxi(\cdot)$ is an unknown function. We
train the model for $\bxi(\cdot)$ using the training data and invert
the function using Bayesian means to estimate the unknown
$\bs^{(new)}$ at which the {test data} is realised.

As discussed above, we vectorise the intrinsically $j\times
k$-dimensional matrix-variate data sets as $jk$-dimensional vectors. In
this treatment, as a measurement is rendered vector-valued,
$\bxi(\cdot)$ is vector-valued and $\bxi(\cdot)$ can be modelled by a
vector-variate ${\cal GP}$ so that realisations from this ${\cal GP}$
are jointly matrix normal. Thus, we consider the $j$ number of
measurements of the $k$-dimensional observable, as a
$jk$-dimensional observed vector ${\bf v}^{(test)}$. This test data is
realised at the unknown value $\bs^{(new)}$ of $\bS$.
Again, a $j\times k$-dimensional synthetic data matrix 
is treated as a $jk$-dimensional synthetic data vector ${\bf
  v}_i$, $i=1,2,\ldots,n$, along the lines of the observed data. Then
all the $n$ synthetic data vectors
together comprise the training data ${\cal D}_s=({\bf v}_1\vdots {\bf
  v}_2\vdots\ldots\vdots{\bf v}_n)^T$ where ${\bf v}_i$
 is generated at the chosen value ${\bs}_i^\star$ of $\bS$, $i=1,\ldots,n$.
Given our treatment
of ${\bf v}_i$ as a $jk$-dimensional vector, the training data set
${\cal D}_s$ is a matrix with $n$ rows and $jk$ columns. 

Thus in this treatment, we have $n$ $jk$-dimensional synthetic data
vectors (inputs), each generated at a chosen value of the model
parameter vector (target), i.e. we have the $n$ observations $({\bf
  v}_1,\bs_1^\star), \ldots, ({\bf v}_n,\bs_n^\star)$, and the aim is
to predict the unknown model parameter vector ${\bs}^{(new)}$ at which
the input is the test data, i.e. the data vector ${\bf
  v}^{(test)}$. In this paradigm of supervised learning akin to the
discussion in \ctn{neal1998}, a predictive distribution of
$\bs^{(new)}$ is sought, conditioned on the test data ${\bf
  v}^{(test)}$ and the training data ${\cal D}_s=({\bf v}_1\vdots{\bf
  v}_2,\vdots\ldots\vdots{\bf v}_n)^T$.

We begin the discussion on the model by elaborating on the detailed
structure of the used ${\cal GP}$. In this section we ignore
measurement errors and present our model of these $n$ vector-variate
functions. Later in Section~\ref{subsec:measerr}, we delineate the
method used to take measurement uncertainties on board.

As the data are vectorised as $jk$-dimensional vectors, 
$\bxi(\cdot)$ is also rendered a $jk$-variate vector function
whose ${\ell}$-th component function is $\xi_\ell(\cdot)$. Then we can write
 ${\bf v}_i=\bxi(\bs_i):=(\xi_1(\bs_i),\ldots,\xi_{jk}(\bs_i))^T$, $\forall
\:i=1,\ldots,n$. We model the $jk$-dimensional function $\bxi(\cdot)$
with a $jk$-dimensional ${\cal GP}$, so that one realisation $\{\bxi(\bs_1),\bxi(\bs_2),\ldots,\bxi(\bs_n)\}$, from
this ${\cal GP}$, is jointly
matrix normal, with adequate parametrisation. We represent this as
\begin{equation}
\{\bxi(\bs_1),\bxi(\bs_2),\ldots,\bxi(\bs_n)\} \sim {\cal MN}_{n,jk}(\bmu, \bA,\bOmega),
\label{eqn:mn}
\end{equation}
where the mean matrix of this matrix normal distribution is the
$n\times jk$-dimensional matrix $\bmu$, the left covariance matrix is
the $n\times n$-dimensional $\bA$ and the right covariance matrix is
the $jk\times jk$-dimensional matrix $\bOmega$. These individual
matrix-variate parameters of this distribution stem from the
parametrisation of the high-dimensional ${\cal GP}$ that is used to
model $\bxi(\cdot)$; we discuss such parametrisation below. Before
proceeding to that, we note that Equation~\ref{eqn:mn} is the same as
saying that the likelihood is matrix normal.
\subsection{Parameters of the matrix-normal distribution}
\label{subsec:detailed}
\noindent
Assuming $\bxi(\cdot)$ to be continuous, the applicability of a
stationary covariance function is expected to suffice. We choose to
implement the popularly used square exponential covariance function
\ctp{rasmussen,scholkopf,Santner03}. This covariance function is easy
to implement and renders the sampled functions smooth and infinitely
differentiable.
Also, we relax the choice of a zero mean
function though that is another popular choice. Instead we choose to
define the mean function in a way that is equivalent to the suggestion
that the data is viewed as centred around a linear model with the
residuals characterised by a vector-variate ${\cal GP}$
\ctp{tony78,cressie}. We then integrate over all such possible global
intercepts to arrive at a result that is more general than if the mean
is fixed at zero.
An advantage of the non-zero mean function is that in the limit of the
smoothness parameters (characterising the smoothness of the functions
sampled from this ${\cal GP}$) approaching large values, the
random function reduces to a linear regression model. This appears
plausible, as distinguished from the result that in this limit
of very large smoothness, the random function will concur with the
errors, as in models with a zero mean function.

The non-zero mean function $\bmu(\cdot)$ of the ${\cal GP}$ is
represented as factored into a matrix $\bH$ that bears information
about its shape and another ($\bB$) that tells us about its amplitude,
or the extent to which this chosen mean function deviates from being
zero. Thus, $\bmu(\cdot):=\bH\bB$, where
\begin{eqnarray}
{\bH}^T &:=& [\bh^{(m\times 1)}(\bs_1),\ldots,\bh^{(m\times 1)}(\bs_n)],\quad\mbox{with}\nonumber \\
m &:=& d+1 \nonumber \\
\bh^{(m\times 1)}(\bs_i)&=&(1,s_i^{(1)},s_i^{(2)},\ldots,s_i^{(d)})^T
\label{eqn:mean}
\end{eqnarray}
where $\bs_i=(s_i^{(1)},s_i^{(2)},\ldots,s_i^{(d)})^T$ for
$i=1,\ldots,n$ and we have recalled the suggestion that such a
non-zero mean function be expressed in terms of a few basis functions
\ctp{rasmussen}, (prompting us to choose to fix this functional form
such that $\bh(\bs):=(1,\bs)^T$ for all values of $\bS$). A similar
construct was used by \ctn{blight} who performed a ${\cal GP}$-based
polynomial regression analysis. Thus, in our treatment, $\bh(\cdot)$
is a $(d+1)$-dimensional vector. The coefficient matrix $\bB$ is
\begin{equation}
\bB=(\bbeta_{11},\ldots,\bbeta_{j1},\ldots,\bbeta_{1k},\ldots,\bbeta_{jk})
\label{eqn:B}
\end{equation}
where for $p=1,\ldots,j$, $p'=1,\ldots,k$, $\bbeta_{pp'}$ is an
$m$-dimensional column vector. As we choose to set $m=d+1$, $\bB$ is a
matrix with $d+1$ rows and $jk$ columns.

The covariance function of the ${\cal GP}$ is again represented as
factored into a matrix $\bOmega$ that tells us about the
amplitude of the covariance and another $\bA$ that bears information
about its shape. The amplitude matrix $\bOmega$ is $jk\times
jk$-dimensional and is defined as
\begin{equation}
\bOmega = \bSigma\otimes\bC
\label{eqn:omega}
\end{equation}
where $\bSigma$ is the $k\times k$ matrix telling us the amplitude of
the covariance amongst the $j$ different observations, for each of the
$k$ components of the data vector, at a fixed value of $\bS$. On the
other hand, $\bC$ is the $j\times j$ matrix giving the amplitude of
covariance amongst the $k$ different components of the vector-valued
observable, at each of the $j$ observations, at a given value of
$\bS$. Thus in our application, an element of $\bSigma$ is the matrix
is the amplitude of the covariance of a given component of the
velocity vectors of the different stars that are observed. This matrix
can then tell us about how a given component of the velocity vectors
of the different stars in the observed sample, correlate with each
other. On the other hand, the matrix $\bC$ informs us about the
amplitude of covariance amongst the different components of the
velocity vectors of a given star in the sample.

We realise that under the assumption of Gaussian
errors in the measurements, the error variance matrix will be added to
$\bOmega$. We discuss this in detail later in
Section~\ref{subsec:measerr}.

The shape of the covariance function is borne by the matrix $\bA$
which is $n\times n$-dimensional. Given our choice of square
exponential covariance function, it is defined as
\begin{eqnarray}
\bA^{(n\times n)} & := & [a(\cdot,\cdot)],
\quad\mbox{where}\nonumber\\ 
a(\bs,\bs')&\equiv&\exp\{-(\bs-\bs')^T\bQ(\bs-\bs')\},\quad
\label{eqn:a}
\end{eqnarray} 
for any 2 values $\bs$ and $\bs'$ of $\bS$. Here, $\bQ^{(d\times d)}$
represents the inverse of the scale length that underlies correlation
between functions at any two values of the function variable. In other
words, $\bQ$ is the matrix of the smoothness parameters. Thus, $\bQ$
is a matrix that bears information about the smoothness of the sampled
functions; it is a diagonal matrix consisting of $d$ non-negative
smoothness parameters denoted by $b_1,\ldots,b_d$. In other words, we
assume the same smoothness for each component function of
$\bxi(\cdot)$. This smoothness is determined by the parameters
$b_1,\ldots,b_d$.  We will learn these smoothness parameters in our
work from the data. Of course, though we say that the smoothness is
learnt in the data, the underlying effect of the choice of the square
exponential covariance function on the smoothness of the sampled
functions is acknowledged. Indeed, as \ctn{snelson} states, one
concern about the square exponential function is that it renders the
functions sampled from it as artificially smooth. An alternative
covariance function, such as the Matern class of covariances
\ctp{matern,gneiting_matern,snelson}, could give rise to sampled
functions that are much rougher than those obtained using the square
exponential covariance function, for the same values of the
hyper-parameters of amplitude and scale that characterise these
covariance functions(see Chapter~1 of Snelson's thesis).

Let $\omega_{r\ell}$ denote the $(r,\ell)$-th element of $\bOmega$,
$c_{r\ell}$ the $(r,\ell)$-th element of $\bC$ and let
$\sigma_{r\ell}$ denote the $(r,\ell)$-th element of $\bSigma$.  Let
the $\ell$-th component function of $\bxi(\cdot)$ be $\xi_\ell(\cdot)$
with $\ell=m_1 k + m_2$, where $\ell=1,\ldots,jk$ and $m_2=1,2,\ldots,k$, 
$m_1=0,1,\ldots,j-1$. Then the correlation between the components
of $\bxi(\cdot)$ yields the following correlation structures:
\begin{align}
corr\left(\xi_{m_1 k+m_2}(\bs_i),\xi_{m'_1 k+ m_2}(\bs_i)\right)&=\frac{\sigma_{m_1 m'_1}}{\sqrt{\sigma_{m_1 m_1}\sigma_{m'_1 m'_1}}} \ {\forall} \ \ m_2, i \ \ \mbox{and}\ \ m_1\neq m'_1\label{eq:corr1}\\
corr\left(\xi_{m_1 k+ m_2}(\bs_i),\xi_{m_1 k+ m'_2}(\bs_i)\right)&=\frac{c_{m_2 m'_2}}{\sqrt{c_{m_2 m_2}c_{m'_2 m'_2}}} \ \forall\:m_1, i \ \ 
\mbox{and}\ \ m_2\neq m'_2\label{eq:corr2}\\
corr\left(\xi_{m_1 k+ m_2}(\bs_i),\xi_{m'_1 k+ m'_2}(\bs_i)\right)&=\frac{c_{m_2 m'_2}\sigma_{m_1 m'_1}}{\sqrt{c_{m_2 m_2}\sigma_{m_1 m_1}c_{m'_2 m'_2}\sigma_{m'_1 m'_1}}} {\forall}i,m_1\neq m'_1,m_2\neq m'_2  \label{eq:corr3}\\
corr\left(\xi_{\ell}(\bs_1),\xi_{\ell}(\bs_2)\right)&=a(\bs_1,\bs_2) {\forall} \ \ \ell \ \ {\mbox{and}}\ \ \bs_1\neq \bs_2\label{eq:corr4}
\end{align}
The 1st of the above 4 equations shows the correlation between the
component functions for the same component of the vector-valued
observable at 2 (of the $j$) different measurements, taken at a given
value of the $\bS$. For a given measurement, the correlation between 2
different components of (the $k$ components of) the observable is
given by the 2nd equation above. For a given value of $\bS$, if we
seek the correlation between the component functions for 2 different
measurements of 2 different components of the observables, this is provided in the 3rd equation. The correlation between component functions for 2 different
values of $\bS$ is given in the last of the above 4 equation. Then
these 4 correlations give the full correlation structure amongst 
components of $\bxi(\cdot)$.
\subsection{Likelihood}
\label{subsec:likelihood}
\noindent
The training data is the $n\times jk$-dimensional matrix ${\cal D}_s=({\bf
  v}_1\vdots {\bf v}_2\vdots\ldots\vdots{\bf v}_n)^T$ where ${\bf
  v}_i$ is the $jk$-dimensional synthetic motion vector generated at
design vector $\bs_i^\star$, $i=1,2,\ldots,n$. To
express the likelihood, we recall that the distribution of the
training data $\{{\bf v}_1,{\bf v}_2,\ldots,{\bf v}_n\}$, i.e. the
joint distribution of $\{\bxi(\bs_1^\star),
\bxi(\bs_2^\star),\ldots,\bxi(\bs_n^\star)\}$ is matrix normal
(Equation~\ref{eqn:mn}). In order to achieve this likelihood, we
rewrite the $\bS$-dependent parameters of this matrix normal
distribution at the values of $\bS$ at which the training data ${\cal
  D}_s$ is realised, i.e. in terms of the design vectors. Thus, we
define
\begin{itemize}
\item the $n\times jk$-dimensional mean function $\bH_D\bB$,
  where the linear form of the mean structure is contained in
  $\bH_D^{(n\times m)}:=[\bh^{(m\times 1)}(\bs_1^\star),\ldots,\bh^{(m\times 1)}(\bs_n^\star)]$ (and the coefficient matrix $\bB$ is defined
  in Equation~\ref{eqn:B}). 
\item the square exponential factor in the covariance matrix $\bA_D^{(n\times n)}:=[\exp\{-(\bs^\star-\bs'^\star)^T\bQ(\bs^\star-\bs'^\star)\}]$ (see Equation~\ref{eqn:a}). 
\end{itemize}
Then it follows from the matrix normal distribution of 
Equation~\ref{eqn:mn}--with mean function defined in
Equation~\ref{eqn:mean} and Equation~\ref{eqn:B}, and covariance
matrix defined using Equation~\ref{eqn:a} and Equation~\ref{eqn:omega}--that
${\cal D}_s$ is distributed as matrix normal with mean matrix $\bH_D\bB$, left
covariance matrix $\bA_D$ and right covariance matrix $\bOmega$, i.e.
\begin{equation}
[{\cal D}_s\mid \bB,\bC,\bSigma,\bQ]\sim{\cal MN}_{n,jk}(\bH_D\bB,\bA_D,\bOmega)
\label{eq:matrix_normal}
\end{equation}
Thus, using known ideas about the matrix normal distribution - see
\ctn{Dawid81}, \ctn{Carvalho07} - we write
\begin{equation}
[{\cal D}_s\mid\bB,\bC,\bSigma,\bQ] =
\frac{1}{(2\pi)^{\frac{njk}{2}}|\bA_D|^{\frac{jk}{2}}|\bOmega|^{\frac{n}{2}}}\exp\left\{-\frac{1}{2}tr\left[\bOmega^{-1}({\cal D}_s-\bH_D\bB)^T\bA^{-1}_D({\cal D}_s-\bH_D\bB)\right]\right\}
\label{eq:matrix_normal_density}
\end{equation}
The interpretation of the above is that the $r$-th row of $[{\cal
    D}_s\vert {\bf B},{\bf\Sigma},{\bf C}, {\bQ}]$ is multivariate
normal with mean corresponding to row of the mean matrix
$\bH_D\bB$ and with covariance matrix $\bOmega$. Rows
$r$ and $\ell$ of $[{\cal D}_s\vert {\bf B},{\bf\Sigma},{\bf
    C},{\bQ}]$ has covariance matrix $a(\bs_r,\bs_{\ell})\bOmega$.
Similarly, the $\ell$-th column of it is distributed as multivariate
normal with mean being the $\ell$-th column of
$\bH_D\bB$ and with covariance matrix
$\omega_{\ell,\ell}\bA_D$, where $\omega_{r,\ell}$ denotes the
$(r,\ell)$-th element of $\bOmega$. The covariance between columns $r$
and $\ell$ is given by the matrix $\omega_{r,\ell}\bA_D$.
\subsection{Estimating ${\bs}^{(new)}$}
\noindent
In order to predict the unknown model parameter vector
${\bs}^{(new)}$ when the input is the measured real data vector
${\bf v}^{(test)}$, we would need the posterior predictive
distribution of ${\bs}^{(new)}$, given ${\bf v}^{(test)}$ and the
training data ${\cal D}_s$. This posterior predictive is usually
computed by integrating over all the matrix-variate ${\cal GP}$
parameters realised at the chosen design vectors
$\bs_1^\star,\ldots,\bs_n^\star$.

While it is possible to analytically integrate over $\bB$ and $\bC$,
$\bSigma$ and $\bQ$ cannot be analytically integrated out. In fact, we
find it useful to learn the $d$ smoothing parameters i.e. the $d$
diagonal elements of ${\bQ}$, given the data. Thus, one useful
advantage of our method is that the smoothness of the process does not
need to be imposed by hand, but can be learnt from the data, if
desired.

Given that we are then learning of $\bs^{(new)}$, $\bSigma$ and $\bQ$,
we rephrase our motivation as seeking to compute the joint posterior
probability of ${\bs}^{(new)}$, $\bQ$ and $\bSigma$, conditional on
the real data and the training data, for a choice of the design
set. In fact, we achieve a closed form expression of this joint
posterior of ${\bs}^{(new)}$, $\bQ$ and $\bSigma$, by integrating over
the other hyper-parameters, namely, the amplitude of the mean function
($\bB$) and the matrix $\bC$ that bears information about covariance
between different components of the data vector for each of the
$j$ observations, at a fixed value of $\bS$. From this closed form
expression, the marginal posterior probability densities of $\bQ$,
$\bSigma$ and any of the $d$ components of the ${\bs}^{(new)}$ vector
can be obtained, using the transformation based MCMC sampling method
\ctp{Duttall} that we adopt.

Thus, for a given choice $\bs_1^\star,\ldots,\bs_n^\star$ of the design
vectors, the posterior distribution $[{\bf s}^{(new)}, {\bSigma},
  {\bQ}\vert {\bf v}^{(test)},{\cal D}_s]$ is sought, by marginalising
$[{\bf s}^{(test)}, {\bf\Sigma}, {\bQ}, {\bB},{\bC}\vert
  {\bf v}^{(test)},{\cal D}_s]$ over the process matrices ${\bB}$ and
${\bC}$. 
\subsection{Priors used}
\label{subsec:marginalized}
\noindent
We use uniform prior on $\bB$ and a simple non-informative prior on
$\bC$, namely, $\pi(\bC)\propto\mid\bC\mid^{-(j+1)/2}$.  As for the
priors on the other parameters, we assume uniform prior on $\bQ$ and
use the non-informative prior
$\pi(\bSigma)\propto\mid\bSigma\mid^{-(k+1)/2}$. The prior information
available in the literature will be considered to select the prior on
$\bs^{(new)}$; below we use uniform priors on all components of the
$\bs^{(new)}$ vector (see Section~\ref{sec:results} for greater
details in regard to the application that we discuss later).
\subsection{Posterior of $\bs^{(new)}$ given training and test data}
\label{subsec:posterior}
\noindent
Since our interest lies in estimating $\bs^{(new)}$, given the
real (test) data and the simulated (training) data, as well as in learning
the smoothness parameter matrix $\bQ$ and the matrix $\bSigma$ that
bears the covariance amongst the $j$ observables, we compute the
joint posterior probability density $[\bs^{(new)},\bQ,\bSigma\mid{\bf
    v}^{(test)},{\cal D}_s]$. As expressed above, we achieve this by
writing $[\bs^{(new)},\bB,\bC,\bQ,\bSigma\mid{\bf v}^{(test)},{\cal
    D}_s]$ and marginalise over $\bB$ and $\bC$.

To construct an expression for this posterior distribution, we first
collate the training and test data to construct the augmented data set
${\cal D}^T_{aug}=({\bf v}_1^T\vdots\ldots\vdots{\bf v}_n^T\vdots({\bf
  v}^{(test)})^T)$. Then the set of values of the model parameter
vector $\bS$ that supports ${\cal D}_{aug}$ is
$\{\bs_1^\star,\ldots,\bs_n^\star,\bs^{(new)}\}$ of which only 
$\bs^{(new)}$ is unknown. 

We next write the $\bS$-dependent matrix-variate parameters at those
values of $\bS$ at which the augmented data set is realised. Thus we
define 
\begin{itemize}
\item $\bH_{{\cal D}_{aug}}^{((n+1)\times m)}:=[\bh^{(m\times
    1)}(\bs_1^\star),\ldots,\bh^{(m\times
    1)}(\bs_n^\star),\bh^{(m\times 1)}(\bs^{(new)})]$, where our
choice of the functional form of $\bh(\cdot)$ has been given in
Section~\ref{sec:modelling} and we also set $m=d+1$,
\item $\bA_{{\cal D}_{aug}}^{((n+1)\times
  (n+1))}:=[\exp\{-(\bs'_i-\bs'_{i'})^T\bQ(\bs'_i-\bs'_{i'})\}]$ where
$\bs'_i$ and $\bs'_{i'}$ are members of the set
$\{\bs_1^\star,\ldots,\bs_n^\star,\bs^{(new)}\}$,
\item $\bM_{aug} := \bA^{-1}_{{\cal D}_{aug}} -\bA^{-1}_{{\cal D}_{aug}}\bH_{{\cal D}_{aug}}[\bH_{{\cal D}_{aug}}^T\bA^{-1}_{{\cal D}_{aug}}\bH_{{\cal D}_{aug}}]^{-1}\bH_{{\cal D}_{aug}}^T\bA^{-1}_{{\cal D}_{aug}}$.
\item $({\cal D}_{aug}^T\bM_{aug}{\cal D}_{aug})^{(jk\times jk)}:=[\bM^*_{tu};
  t,u=1,\ldots,k]$, where $\bM^*_{tu}$ is a matrix with $j$ rows and
$j$ columns. Given $\bSigma$, we define $m=d+1$ and $\psi^{-1}_{tu}$
as the $(t,u)$-th element of $\bSigma^{-1}$, so that
$(n+1-m)k{\hat{\bC}}_{GLS,aug}:=\sum_{t=1}^k\sum_{u=1}^k\psi^{-1}_{tu}\bM^*_{tu}$, where $(n+1-m)k{\hat{\bC}}_{GLS,aug}$ is used in the closed-form expression for $[\bs^{(new)},\bQ,\bSigma\mid{\bf v}^{(test)},{\cal D}_s]$ that we seek.
\end{itemize}

The priors used on $\bB$, $\bC$, $\bQ$, $\bSigma$ and $\bs^{(new)}$
are listed in Section~\ref{subsec:marginalized}.  Using
these, and recalling Equation~\ref{eq:matrix_normal_density}, we get the joint posterior probability density of all unknown parameters given all data, i.e. \\[4mm]
$[\bs^{(new)},\bQ,\bB,\bSigma,\bC\mid{\bf v}^{(test)},{\cal
    D}_s]\propto [{\cal
    D}_{aug}\mid\bB,\bSigma,\bC,\bQ,\bs^{(new)}][\bB,\bSigma,\bC,\bQ,\bs^{(new)}],$ \\[4mm]
which we then marginalise 
over $\bB$ and $\bC$ to get the joint posterior 
$[\bs^{(new)},\bQ,\bSigma\mid{\bf v}^{(test)},{\cal D}_s]$, as\\[4mm]

$[\bs^{(new)},\bQ,\bSigma\mid{\bf v}^{(test)},{\cal D}_s]$
\begin{eqnarray}
&=&\int\int[\bs^{(new)},\bQ,\bB,\bSigma,\bC\mid{\bf v}^{(test)},{\cal D}_s]d\bB d\bC\nonumber\\
&\propto&|\bA_{{\cal D}_{aug}}|^{-\frac{jk}{2}}
|\{\bH_{{\cal D}_{aug}}\}^T\{\bA_{{\cal D}_{aug}}\}^{-1}\{\bH_{{\cal D}_{aug}}|^{-\frac{jk}{2}}\times|\bSigma|^{-\frac{j(n+1-m)+k+1}{2}}|(n+1-m)k\hat\bC_{GLS,aug}|^{-\frac{(n+1-m)k}{2}} \nonumber \\
\label{eq:int2}
\end{eqnarray}
Thus, we obtain a closed-form expression of the joint posterior of
$\bs^{(new)},\bQ,\bSigma$, given training and test data, for a given
choice of the design matrix (Equation~\ref{eq:int2}), up to a
normalising constant. The ${\cal GP}$ prior is strengthened by the $n$
number of samples taken from it at the training stage. We sample
from the achieved posterior using MCMC techniques to achieve the
marginal posterior probabilities of $\bQ$, $\bSigma$ or any component
of ${\bs}^{(new)}$, given all data. 
We conduct posterior inference using the TMCMC methodology
\ctp{Duttall} that works by constructing proposals that are
deterministic bijective transformations of a random
vector drawn from a chosen distribution. 
\subsection{Errors in measurement}
\label{subsec:measerr}
\noindent
In our application, the errors in the measurements are small and will
be ignored for the rest of the analysis. In general, when errors in
the measurements that comprise the training data and the test data are
not negligible, we assume Gaussian measurement errors
${\bf\varepsilon}_t$, in ${\bf v}_t$, with $t=1,2,\ldots$, such that
${\bf\varepsilon}_t\sim{\cal N}_{jk}(\bzero,{\bf\varsigma})$, where
$\varsigma=\bSigma_1\otimes\bSigma_2$; $\bSigma_1,\bSigma_2$ being
positive definite matrices. If both $\bSigma_1$ and $\bSigma_2$ are
chosen to be diagonal matrices, then $\varsigma$ is a diagonal matrix;
assuming same diagonal elements would simplify $\varsigma$ to be of
the form $\varphi\times\bI$, where $\bI$ is the $jk\times jk$-th order
identity matrix. This error variance matrix $\varsigma$ must be added
to $\bOmega$ before proceeding to the subsequent calculations. TMCMC
can be then be used to update $\varsigma$.
\section{Case study}
\label{sec:casestudy}
\noindent
Using the methodology discussed above we attempt an estimate of the
unknown Milky Way feature parameter vector ${\bS}\in{\mathbb R}^d$
using the available stellar velocity data. In our application,
the dimensionality of ${\bS}$ is 2 as we estimate the coordinates of
the radial location $r_\odot$ of the Sun with respect to the Galactic
centre and the angular separation $\phi_\odot$ of the Sun-Galactic
centre line from a pre-set line in the Milky Way disk (see Figure~1 in
supplementary section {\bf{S-1}}). Then for the Sun, $R=r_\odot$ and
$\Phi=\phi_\odot$ where the variable $R$ gives radial distance from
the Galactic centre of any point on the disk of the Milky Way and the
variable $\Phi$ gives the angular separation of this point from this
chosen pre-set line. The reason for restricting our application to the
case of $d$=2 is the existence of simulated stellar velocity data (aka
training data) generated by scanning over chosen guesses for $r_\odot$
and $\phi_\odot$, with all other feature parameters held constant. If
simulated data distinguished by choices of other Milky Way feature
parameters become available, then the implementation of such data as
training data will be possible, allowing then for the learning of
Milky way parameters in addition to $r_\odot$ and $\phi_\odot$. In
this method, computational costs are the only concern in extending to
cases of $d>2$; extending to a higher dimensional ${\bS}$ only
linearly scales computational costs (Section~\ref{sec:discussions}).


Also, the stellar velocity vector is 2-dimensional, i.e. $k$=2 in this
application. Then the measured data in this application is a $j\times
2$-dimensional matrix. In our Bayesian approach, a much smaller $j$
(=50) allows for inference on the unknown value ${\bs}^{(new)}$ of the
Milky Way feature parameter vector, than $j\sim$3000 that is demanded
by the aforementioned calibration approach used by
\ctn{chakrabarty07}.

In our application, the available data include the measured or test
data and 4 sets of synthetic (or training) data sets obtained via
dynamical simulations of each of 4 distinct base-astronomical models
of our galaxy, advanced by \ctn{chakrabarty07}.  As the analysis is
performed with each training data set at a time, we do not include
reference to the corresponding base model in the used notation. The
simulated data presented in \ctn{chakrabarty07} that we use here, is
generated at 216 distinct values of ${\bS}$, i.e. $n$=216.  Thus, our
design set comprises the 216 chosen values of $\bS$:
$\bs_1^\star,\ldots,\bs_{216}^\star$. For each of the 4 base
astrophysical models, at each chosen ${\bs}_i^\star$, 50 2-dimensional
stellar velocity vectors are generated from dynamical simulations of
that astrophysical model (of the Milky Way), performed at that value
of ${\bS}$. These 50 2-dimensional velocity vectors are treated in our
work as a 50$\times$2=100-dimensional motion vector ${\bf v}_i$;
$i=1,\ldots,216$. Then at the 216 {design vectors},
$\bs_1^\star,\ldots,\bs_{216}^\star$, 216 motion vectors are
generated: ${\bf v}_1,\ldots,{\bf v}_{216}$.  Then the training data
in our work comprises all such motion vectors and is represented as
${\cal D}_s^{(216\times 100)}$. The real or test data is treated in
our work as the 100-dimensional motion vector ${\bf v}^{(test)}$.

As said above, there are 4 distinct training data sets available
from using the 4 base astronomical models of the Milky Way, as considered by
\ctn{chakrabarty07}. 
The choice of the base astrophysical model is distinguished by the
ratio of the rates of rotation of the spiral to the bar,
$\Omega_s/\Omega_b$. That this ratio is relevant to stellar motions in
the Galaxy is due to the fact that $\Omega_s/\Omega_b$ can crucially
control the degree of chaos in the Galactic model\footnotemark.
\footnotetext{For example, it is well known in chaos theory that when
  $\Omega_s/\Omega_b$ is such that one of the radii at which the bar
  and the stellar disk resonate, concurs with a radius at which the
  spiral and the stellar disk resonate, global chaos is set up in the
  system \ctp{fordwalker}. \ctn{chaksid08} have corroborated that the
  degree of chaos is maximal in the astrophysical Galactic model
  marked by such a ratio ($\Omega_s/\Omega_b$=22/55). They report that
  in models marked by slightly lower ($\Omega_s/\Omega_b$=18/55) or
  higher ($\Omega_s/\Omega_b$ = 25/55) values of this ratio, chaos is
  still substantial. In the Galactic model that precludes the spiral
  however, chaos was quantified to be minimal. It is these 4 states of
  chaos - driven by the 4 values of $\Omega_s/\Omega_b$ - that mark
  the 4 astrophysical models as distinct.}  
Thus, the 4 base
astrophysical models are differently chaotic.  This results in 4
distinct simulated velocity data sets ${\cal D}_s^{(1)}$,
${\cal D}_s^{(2)}$, ${\cal D}_s^{(3)}$, ${\cal D}_s^{(4)}$ that bear
the effects of such varying degrees of chaos, each generated at the
chosen design set $\{{\bs}_1^\star,\ldots,{\bs}_n^\star\}$. 
Details of the dynamical simulations
performed on the 4 astrophysical models are given in the supplementary
section {\bf{S-2}}.
\subsection{Details of our implementation of TMCMC}
\label{subsec:implementtmcmc}
\noindent
As indicated above, we use the Transformation-based MCMC (TMCMC)
advanced by\\ \ctn{Duttall} to conduct posterior inference. In TMCMC,
high-dimensional parameter spaces are explored by constructing
bijective deterministic transformations of a low-dimensional random
vector. The random vector of which a proposal density is a
transformation of, can be chosen to be of dimensionality between 1 and
the dimensionality of the parameters under the target posterior. The
acceptance ratio in TMCMC does not depend upon the distribution of the
chosen random vector.  In our application we use TMCMC to update the
entire block $(\bs^{(new)},\bQ,\bSigma)$ at the same time using
additive transformations of a one-dimensional random variable
$\epsilon\sim {\cal N}(0,1)I_{\{\epsilon>0\}}$.  In the $t$-th
iteration, the state of the unknown parameters is
$(\bs^{(new,t)},\bQ^{(t)},\bSigma^{(t)}):=\bvarphi^{(t)}$. We update
$\bvarphi^{(t)}$ by setting, with probabilities $\pi_j$ and
$(1-\pi_j)$, $\varphi^{(t+1)}_j=\varphi^{(t)}_j\pm c_j\epsilon$
(forward transformation) and
${\bvarphi}^{(t+1)}_j={\bvarphi}^{(t)}_j-c_j\epsilon$ (backward
transformation), respectively, where, for $j=1,\ldots,d$, $\pi_j$ are
appropriately chosen probabilities and $c_j$ are appropriately chosen
scaling factors.  Assume that for $j_1\in\mU$,
${\bvarphi}^{(t)}_{j_1}$ gets the positive transformation, while for
$j_2\in\mU^c$, ${\bvarphi}^{(t)}_{j_2}$ gets the backward
transformation. Here $\mU\cup\mU^c=\{1,\ldots,d^*\}$, where
$d^*=2d+\frac{k(k+1)}{2}$. The proposal $\bvarphi^{(t+1)}$ is accepted
with acceptance probability given in Supplementary Section~{\bf S-3}.
Once the proposal mechanism and the initial values are decided,
we discard the first 100,000 iterations of our final TMCMC run
as burn-in and stored the next 1,000,000 iterations for inference. For
each model it took approximately 6 hours on a laptop to generate
1,100,000 TMCMC iterations.
\section{Results using real data}
\label{sec:results}
\noindent 
The training data that we use was obtained by
\ctn{chakrabarty07}, by choosing the solar radial location from the
interval $[1.7,2.3]$ in model units. This explains the motivation for selecting the bounds on $r_\odot$ to be the edges of this interval.
Here, values of distances are expressed in the units implemented in
the base astrophysical models of the Milky Way. However, to make sense
of the results we have obtained, these model units will need to be
scaled to provide values in real astronomical units of distances
inside galaxies, such as the ``kiloparsec'' (abbreviated as
``kpc''). A distance of 1 in model unit scales to
$\displaystyle{\frac{\cal{R}}{{\hat{r}}_\odot}}$kpc, where ${\cal R}$
is the solar radius obtained in independent astronomical studies
\ctp[${\cal R}$=8kpc]{BM} and ${{\hat{r}}_\odot}$ is the estimate of
  the solar radius in our work. The ulterior aim in estimating the
  solar radius is in estimating
  the rotational frequency $\Omega_b$ of the bar, where
  ${\Omega_b} =
  \displaystyle{\frac{v_0}{\frac{\cal{R}}{{\hat{r}}_\odot}}}$, with
  $v_0$=220kms$^{-1}$ and ${\cal{ R}}$=8kpc. Then, we get
  ${\Omega_b} =
  \displaystyle{\frac{220}{\frac{8}{{\hat{r}}_\odot}}}$kms$^{-1}$/kpc.
  See Section~{\bf{S-1}} of the attached supplementary
  material to see a schematic representation of the central bar in the
  Galaxy and Section~{\bf{S-2}} for details of the scaling between
  the model units and real astronomical units.


Our other estimate is of the angular separation between the long axis
of the bar and the line that joins the Sun to the Galactic centre. It is
suggested in past astronomical modelling work to be an acute angle
\ctp{chakrabarty07,hydro,fux}. Indeed, the training data used here was
generated in simulations performed by \ctn{chakrabarty07}, in which
$\phi_\odot$ is chosen from the interval $[0,90^\circ]$. This
motivates the consideration of the interval of $[0,90^{\circ}]$ for
the angular location of the Sun.


Given the bounds on $r_\odot$ and $\phi_\odot$ presented above, in our
TMCMC algorithm, we reject those moves that suggest $r_\odot$ and
$\phi_\odot$ values that fall outside these presented intervals.

The 4 astrophysical models of the Galaxy that were used to generate the
4 training data sets, are marked by the same choice of the value of
$\Omega_b$ and the background Galactic model parameters, while they
are distinguished by the varying choices of the ratio
$\Omega_s:\Omega_b$, where the Galactic spiral pattern rotates with
rate $\Omega_s$. In fact, the astrophysical model $bar\_6$ is the
only one that does not include the influence of the spiral pattern
while the other three astrophysical models include the influence of
both the bar and the spiral. For the astrophysical models
$sp3bar3\_18$, $sp3bar3$ and $sp3bar3\_25$, $\Omega_s:\Omega_b$ is
respectively set to
$18\Omega_b/55,\:22\Omega_b/55,\:25\Omega_b/55$. The physical effect
of this choice is to induce varying levels of chaoticity in the 4
astrophysical models. Thus, \ctn{chaksid08} confirmed that of the 4
models, $bar\_6$ manifests very low chaoticity while $sp3bar3$
manifests maximal chaos, though both $sp3bar3\_18$,
$sp3bar3\_25$ are comparably chaotic.

Ancillary real data needs to be brought in to judge the relative fit
amongst the astrophysical base models. In fact, \ctn{chakrabarty07}
brought in extra information to perform model selection. Such
information was about the observed variance of the components of
stellar velocities and this was used to rule out the model $bar\_6$ as
physically viable, though the other three models were all acceptable
from the point of view of such ancillary observations that are
available. This led to the inference that $\Omega_s\in[18\Omega_b/55,
  25\Omega_b/55]$. 

It is to be noted that if there was 1 data set and we were trying to
fit 4 different models to that same data, then it is very much possible
that for this 1 data set, the average of 4 models could have been
achieved. However, here we are dealing with 4 base models, each of
which is giving rise to a distinct training data set, in fact under
mutually contradicting physics. Therefore, such model averaging is not
relevant for this work. Cross-validation of these 4 models is indeed
possible and we present this in Section~{\bf S-5} of the attached Supplementary Materials.

\begin{figure}
\centering
\label{fig:radius_model1}
\includegraphics[width=4cm,height=4cm]{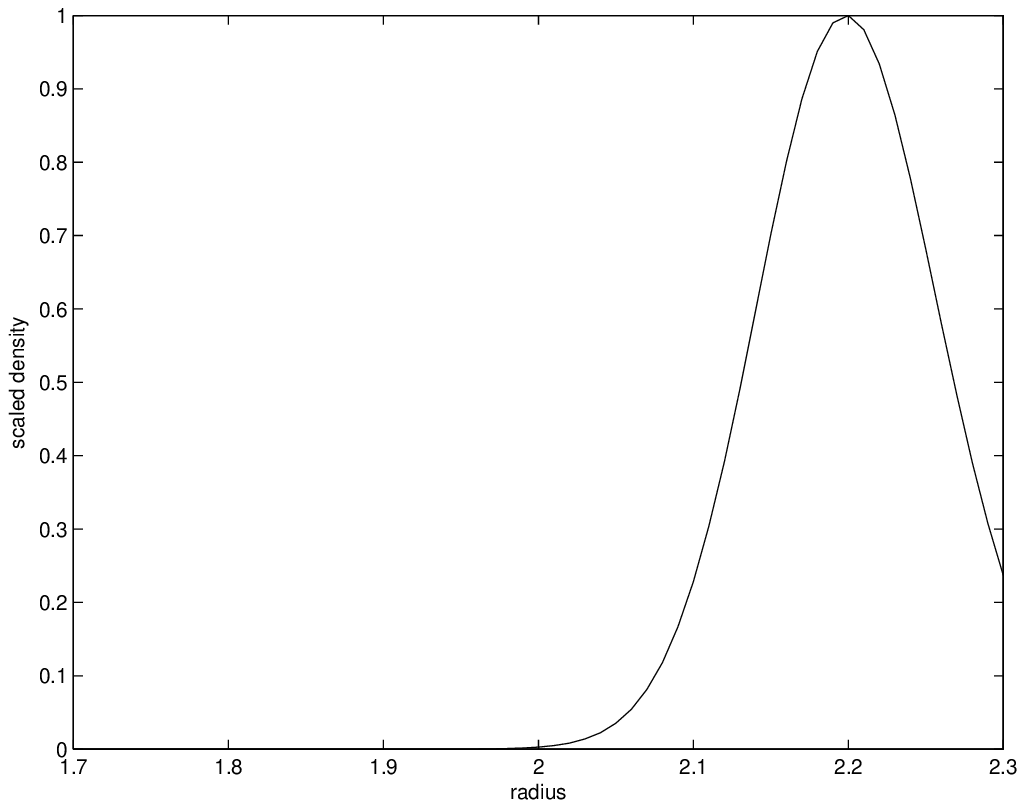}
\hspace{2mm}
\label{fig:theta_model1} 
\includegraphics[width=4cm,height=4cm]{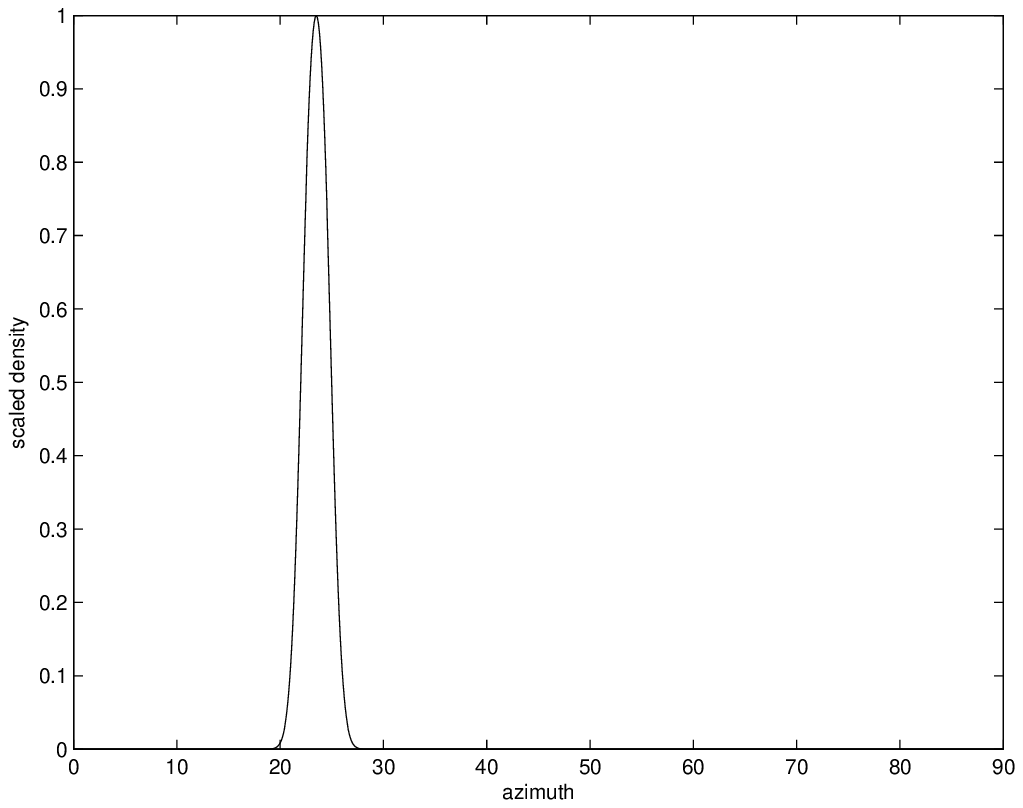}
\caption{Posteriors of $r_\odot$ in model units of $r_{CR}$ and $\phi_\odot$ (in degrees) for the model $bar\_6$.}
\label{fig:model1}
\end{figure}

\begin{figure}
\centering
\label{fig:radius_model2}
\includegraphics[width=4cm,height=4cm]{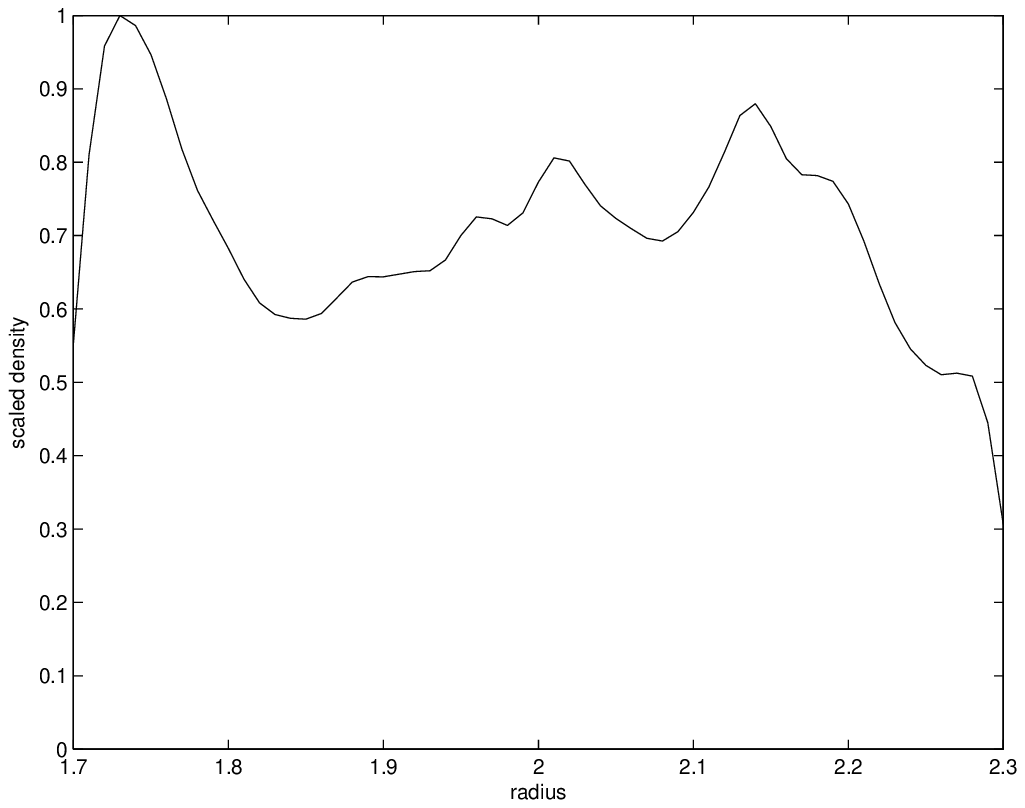}
\hspace{2mm}
\label{fig:theta_model2} 
\includegraphics[width=4cm,height=4cm]{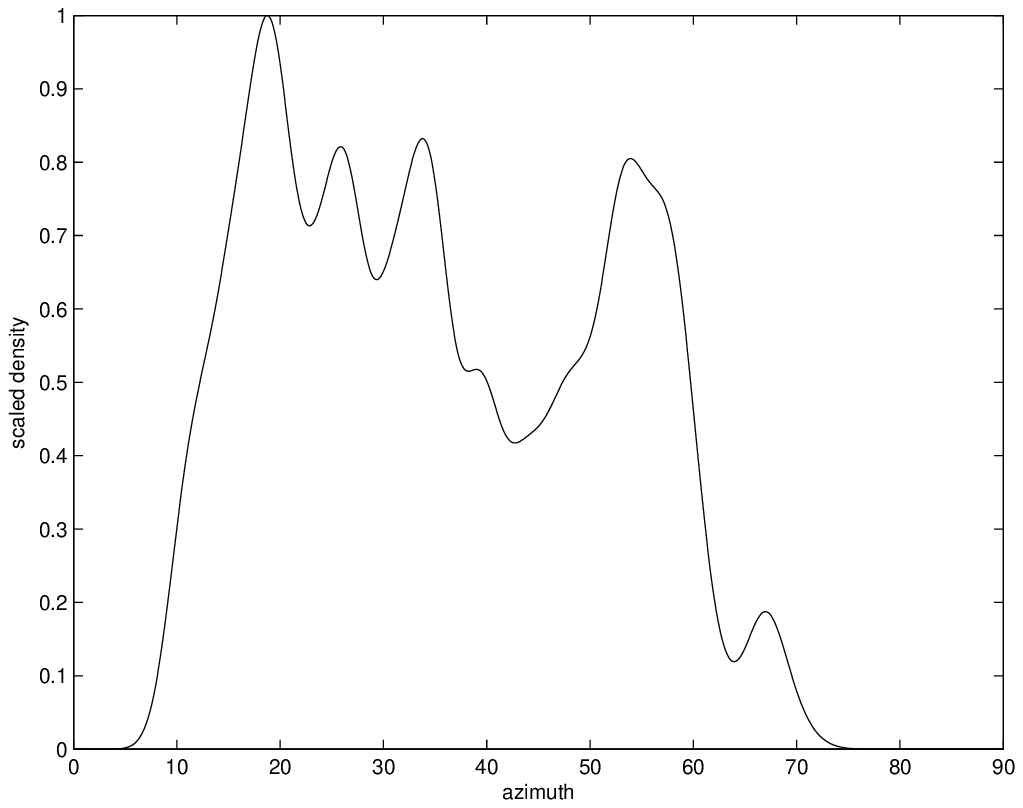}
\caption{Posteriors of $r_\odot$ in units of $r_{CR}$ and $\phi_\odot$ (in degrees) for the model $sp3bar3$.}
\label{fig:model2}
\end{figure}

\begin{figure}[!t]
\centering
\label{fig:radius_model3}
\includegraphics[width=4cm,height=4cm]{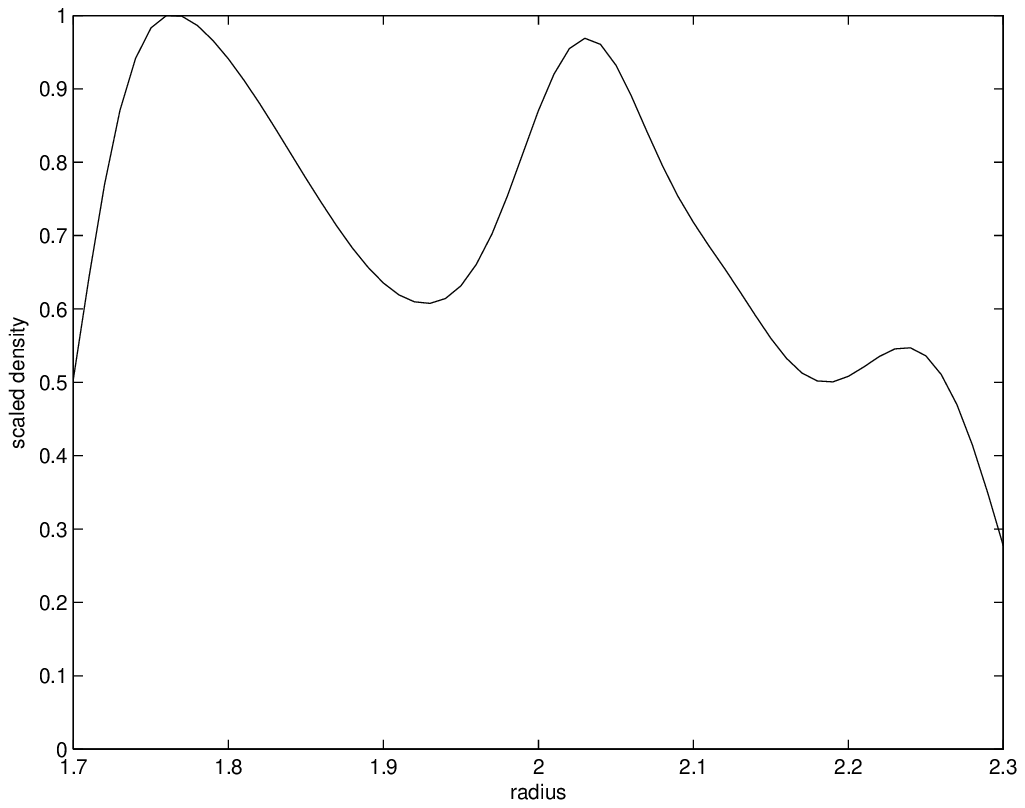}
\hspace{2mm}
\label{fig:theta_model3} 
\includegraphics[width=4cm,height=4cm]{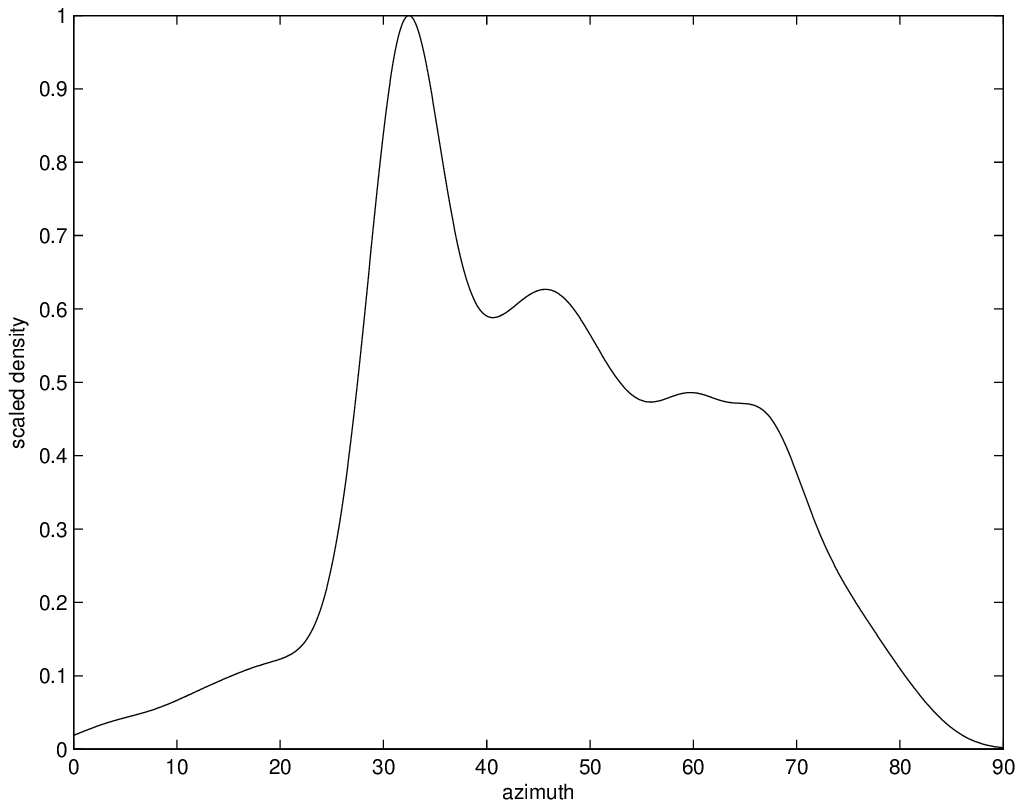}
\caption{Posteriors of $r_\odot$ in model units of $r_{CR}$ and $\phi_\odot$ (in degrees) for the model $sp3bar3\_18$.}
\label{fig:model3}
\end{figure}

\begin{figure}[!ht]
\centering
\label{fig:radius_model4}
\includegraphics[width=4cm,height=4cm]{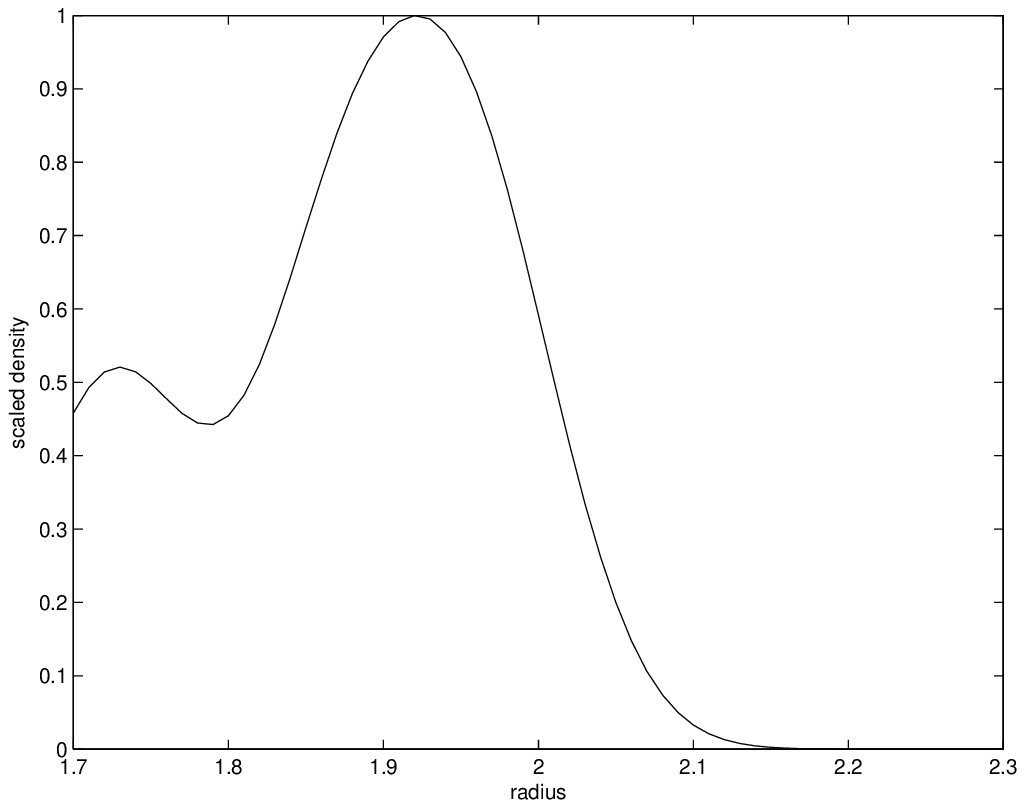}
\hspace{2mm}
\label{fig:theta_model4} 
\includegraphics[width=4cm,height=4cm]{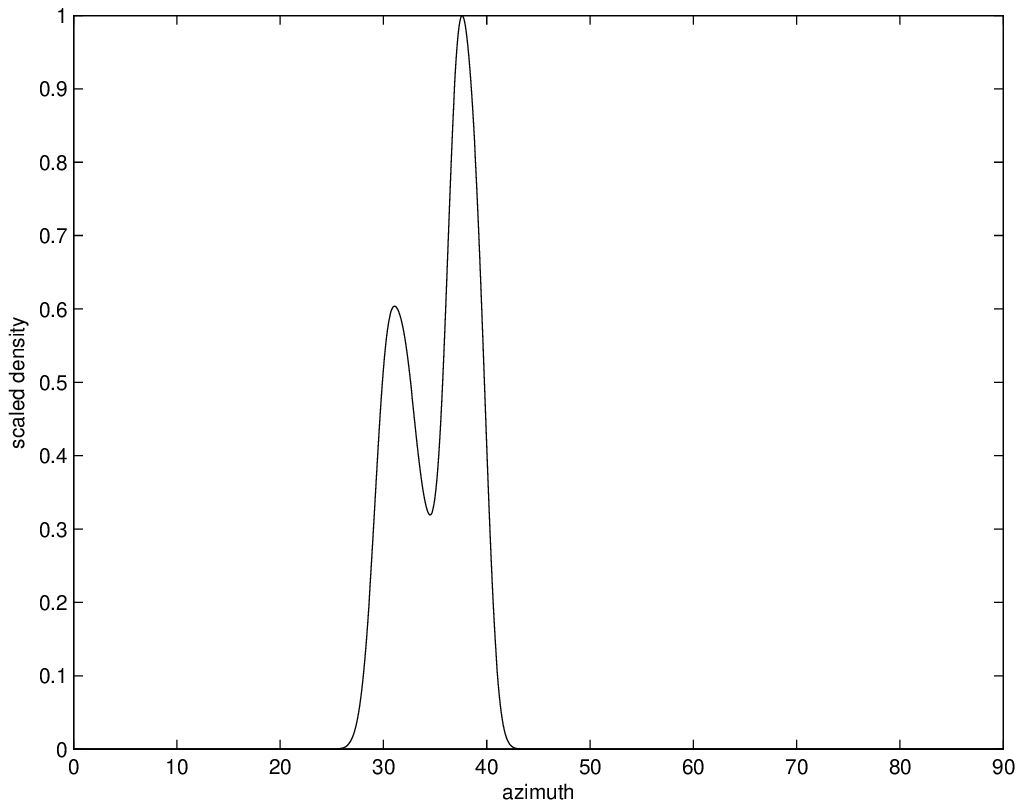}
\caption{Posteriors of $r_\odot$ in units of $r_{CR}$ and $\phi_\odot$ (in degrees) for the model $sp3bar3\_25$.}
\label{fig:model4}
\end{figure}

The marginal posterior densities of $(r_\odot,\phi_\odot)$
corresponding to the 4 base astrophysical models of the Milky Way, are
shown in Figures \ref{fig:model1}, \ref{fig:model2}, \ref{fig:model3}
and \ref{fig:model4}. It merits mention that the multi-modality
manifest in the marginal posterior distributions in 3 of the 4 base
models is not an artifact of inadequate convergence but is a
direct fallout of the marked amount of chaoticity in all 3 base models
except in the model $bar\_6$, \ctp{chaksid08}.
In Section~{\bf S-6}, we discuss the connection between chaos and
consistency of multiple observer locations with available stellar
velocity data.

Table~\ref{table:radius2} presents the posterior mode, the 95\%
highest posterior density (HPD) credible region of $r_\odot$ and
$\phi_\odot$ respectively, associated with the four base models. Here
$r_\odot$ is expressed in the model units of length, i.e. in units of
$r_{CR}$. $\phi_\odot$ is expressed in degrees.  The HPDs are computed
using the methodology discussed in \ctn{Carlin96}. Disjoint HPD
regions, characterise the highly multi-modal posterior distributions
of the unknown location. Using the 95$\%$ HPDs of the estimate
$\hat{r_\odot}$ expressed in model units, and using the independently
known astronomical measurement of the solar radial location as 8kpc,
the bar rotational frequency $\Omega_b$ is computed (see third
enumerated point discussed above) in Table~\ref{table:radius2}.

\begin{table}
\caption{Summary of the posterior distributions of the radial
  component $r_\odot$ and azimuthal component $\phi_\odot$ of the unknown observer location vector
  for the 4 base astrophysical models and the unknown
  bar rotational frequency $\Omega_b$ computed using the 95$\%$ HPDs on the learnt radial location $r_\odot$ in these models.}
 \label{table:radius2}
\begin{center}
{\footnotesize
\begin{tabular}{|c||c|c||c||c|c|}\hline
\multicolumn{1}{|c||}{Model} & \multicolumn{2}{|c||}{$r_\odot$ (in units of $r_{CR}$)} & \multicolumn{1}{|c||}{$\Omega_b$ (in kms$^{-1}$/kpc)} & \multicolumn{2}{|c|}{$\phi_\odot$}\\
\hline
& Mode & 95\% HPD & 95$\%$ HPD & Mode & 95$\%$ HPD\\
\hline
$bar6$ & $2.20$ & $[2.04, 2.30]$ & $[56.1, 63.25]$ &$23.50$ & $[21.20, 25.80]$ \\
$sp3bar3$ & $1.73$ & $[1.70, 2.26]\cup[2.27, 2.28]$ & $[46.75, 62.15]\cup[62.45, 62.7]$ & $18.8$ & $[9.6,61.5]$ \\
$sp3bar3\_18$ & $1.76$ & $[1.70,2.29]$ & $[46.75, 62.98]$ & $32.5$ & $[17.60,79.90]$\\
$sp3bar3\_25$  & $1.95$ & $[1.70,2.15]$ & $[46.75,59.12]$ & $37.6$ & $[28.80,40.40]$ \\
\hline
\end{tabular}
}
\vspace{-0.1in}
\end{center}
\end{table}

Summaries of the posteriors (mean, variance and 95\% credible
interval) of the smoothness parameters $b_1,b_2$ and $\bSigma$ are
presented in Tables \ref{table:smoothness},
\ref{table:sigma_first_diagonal}.  Notable in all these tables are the
small posterior variances of the quantities in question; this is
indicative of the fact that the data sets we used, in spite of the
relatively smaller size compared to the astronomically large data sets
used in the previous approaches in the literature, are very much
informative, given our vector-variate ${\cal GP}$-based Bayesian
approach. Owing to our Gaussian Process approach, the posterior of
$\bSigma$ should be close to the null matrix {\it a posteriori} if the
choice of the design set and the number of design points are
adequate. Quite encouragingly, Table \ref{table:sigma_first_diagonal},
shows that indeed $\bSigma$ is close to the null matrix {\it a
  posteriori}, for all the four models, signifying that the unknown
velocity function has been learned well in all the cases.

\begin{table}
\caption{
Summary of the posterior distributions of the smoothness parameters $b_1,b_2$
for the 4 models.
} 
 \label{table:smoothness}
\begin{center}
{\footnotesize
\begin{tabular}{|c||ccc||ccc|}\hline
\multicolumn{1}{|c||}{Model} & \multicolumn{3}{|c||}{$b_1$}
 & \multicolumn{3}{|c|}{$b_2$}\\
\hline
& Mean & Var & 95\% CI & Mean & Var & 95\% CI \\
\hline
$bar\_6$ & $0.9598155$ & $3.15\times 10^{-9}$ & $[0.959703,0.959879]$ & 1.005078 & $2.85\times 10^{-9}$ & $[1.004985,1.005142]$\\
$sp3bar3$ & $0.8739616$ & $6.72\times 10^{-7}$ & $[0.872347,0.875052]$ & $1.003729$ & $8.98\times 10^{-7}$ & $[1.002500,1.005500]$\\
$sp3bar3\_18$ & $0.9410686$ & $1.46\times 10^{-5}$ & $[0.938852,0.955264]$ & $0.999010$ & $4.08\times 10^{-6}$ & $[0.997219,1.004945 ]$\\
$sp3bar3\_25$  & $0.7597931$ & $5.64\times 10^{-10}$ & $[0.759743,0.759833]$ & $0.992174$ & $2.89\times 10^{-9}$ & $[0.992067,0.992246 ]$\\
\hline
\end{tabular}
}
\vspace{-0.1in}
\end{center}
\end{table}

\begin{table}
\caption{ Summary of the posterior distribution of the diagonal
 and one non-diagonal element of $\bSigma$, from the 4 base astrophysical models.  }
 \label{table:sigma_first_diagonal}
\begin{center}
{\footnotesize
\begin{tabular}{|c||c||c||c|}\hline
\multicolumn{1}{|c||}{Model} & \multicolumn{1}{|c||}{$\sigma_{11}$} & \multicolumn{1}{|c||}{$\sigma_{22}$} & \multicolumn{1}{|c|}{$\sigma_{12}$}\\
\hline
& 95\% CI & 95\% CI & 95\% CI \\
\hline
$bar\_6$ & $[5.40\times 10^{-5}, 4.0\times 10^{-4} ]$ & $[6.20\times 10^{-5},4.76\times 10^{-4}]$ & $[0, 1.30\times 10^{-5}]$\\
$sp3bar3$  & $[3.66\times 10^{-3},1.03\times 10^{-2}]$ & $[6.53\times 10^{-3},1.83\times 10^{-2}]$ & $[-6.40\times 10^{-5}, 2.68\times 10^{-4}]$\\
$sp3bar3\_18$  & $[1.45\times 10^{-3},1.68\times 10^{-1}]$ & $[1.29\times 10^{-3},1.50\times 10^{-1}]$  &  $[ -1.19\times 10^{-4},2.16\times 10^{-3} ]$\\ 
$sp3bar3\_25$  & $[1.21\times 10^{-4},5.69\times 10^{-4}]$ &  $[1.13\times 10^{-4},5.21\times 10^{-4}]$ & $[-1.00\times 10^{-6},1.50\times 10^{-5}]$\\
\hline
\end{tabular}
}
\vspace{-0.1in}
\end{center}
\end{table}
\subsection{Comparison with results in astrophysical literature}
\label{sec:compare}
\noindent
The estimates of the anglar separation of the long axis of the bar
from the Sun-Galactic centre line and the rotation rate of the bar
compare favourably with results obtained by \ctn{chakrabarty07},
\ctn{hydro}, \ctn{debattista}, \ctn{benjamin},\ctn{barbara_11}. A
salient feature of our implementation is the vastly smaller data set
that we needed to invoke than any of the methods reported in the
astronomical literature, in order to achieve the learning of the
two-dimensional vector ${\bS}$ - in fact while in the calibration
approach of \ctn{chakrabarty07}, the required sample size is of the
order of 3,500, in our work, this number is 50. Thus, data sufficiency
issues, when a concern, are well tackled by our method.

Upon the analyses of the viable astrophysical models of the Galaxy,
\ctn{chakrabarty07} reported the result that $r_\odot\in[1.9375,2.21]$
in model units while $\phi_\odot\in[0^\circ,30^\circ]$, where these
ranges correspond to the presented uncertainties on the estimates,
which were however, rather unsatisfactorally achieved (see
Section~\ref{sec:modelling}). The values of the components of ${\bS}$,
learnt in our work, overlap well with these results. As mentioned
above, the models $sp3bar3\_18$, $sp3bar3$ and $sp3bar3\_25$ are
distinguished by distinct values of the ratios of the rotational rates
of the spiral pattern $\Omega_s$ to that of the bar ($\Omega_b$) in
the Galaxy. Then the derived estimate for $\Omega_b$
(Table~\ref{table:radius2}) suggests values of $\Omega_s$ of the Milky
Way spiral.

Another point that merits mentions is that the estimates of $r_\odot$
and $\phi_\odot$ presented by \ctn{chakrabarty07} exclude the model
$sp3bar3$ which could not be used to yield estimates given the highly
scattered nature of the corresponding $p$-value
distribution. Likewise, in our work, the same model manifests maximal
multi-modality amongst the others, but importantly, our approach allows
for the representation of the full posterior density using which, the
computation of the 95$\%$ HPDs is performed.

That the new method is able to work with smaller velocity data sets,
is an important benefit, particularly in extending the application to
galaxies other than our own, in which small numbers of individual
stars are going to be tracked in the very near future for their
velocities, under observational programmes such as PANStarrs
\ctp{panstarrs} and GAIA
\ctp[{\url{http://www.rssd.esa.int/index.php?project=GAIA&page=index}}]{gaia_lindegren,kucinskas};
the sample sizes of measured stellar velocity vectors in these
programmes will be much smaller in external galaxies than what has
been possible in our own. At the same time, our method is advanced as
a template for the analysis of the stellar velocity data that is
available for the Milky Way, with the aim of learning a
high-dimensional Galactic parameter vector; by extending the scope of
the dynamical simulations of the Galaxy, performed on different
astrophysical models of the Milky Way, the Milky Way models will be
better constrained. The mission GAIA - a mission of the European Space
Agency - is set to provide large sets of stellar velocity data all
over the Milky Way. Our method, in conjunction with astrophysical
models, can allow for fast learning of local and global model
parameters of the Galaxy.
\section{Model fitting}
\label{sec:modelfit}
\noindent
In this section we compare the test data with predictions for the
observable that we make at a summary $\tilde{\bs}$ of the posterior of
the model parameter vector $\bS$. To achieve this, we first need to
provide a suitable estimator of the function $\bxi(\cdot)$ that
defines the relatioship between the observable and the model parameter
$\bS$. We attempt to write the conditional distribution of
$\bxi(\tilde{\bs})$ given the augmented data ${\cal D}_a$ that
comprises training data ${\cal D}_s$, augmented by test data
$v^{(test)}$.  Here we consider the test data $v^{(test)}$ realised at
$\bS=\tilde{\bs}$, where we use different candidates for $\tilde{\bs}$.
In particular, we choose $\tilde{\bs}$ to be (1) the median
$\bs^{(median)}$ of the posterior of $\bS$ given ${\cal D}_a$, (2) the mode
$\bs^{(mode)}$ of this posterior, (3) or $\bs^{(u)}$, $u$=1,2,3,4--the end
points of the disjoint 95\% HPD region of the posterior of $\bS$ (see
Table~\ref{table:radius2}).

Since
$\{\bxi(\bs_1),\ldots,\bxi(\bs_n),\bxi(\tilde{\bs})\}$
is jointly matrix-normal,
  $[\bxi(\tilde{\bs})\vert\bxi(\bs_1),\ldots,\bxi(\bs_n))]
 \equiv [\bxi(\tilde{\bs})\vert{\cal D}_s]$, is $jk$-variate normal. The mean function of this multivariate normal, at different $\tilde{\bs}$, is then compared to the test
  data. Thus, the estimate of the function that we seek is ${\mathbb
    E}[\bxi(\bS)\vert {\cal D}_s, \bS, \bQ]$, given the dependence of
  $\bxi(\cdot)$ on the smoothness parameters (elements of $\bQ$) that
  we anticipate.

However, we only know the conditional
of $\bxi(\cdot)$ on all the ${\cal GP}$ parameters, including the ones
that we do not learn from the data, namely $\bB$ and $\bC$. So we need
to marginalise $[\bxi(\cdot)\mid \bSigma,\bB,
  \bC,\bQ,{\cal D}_s]$ over $\bB$ and
$\bC$. To achieve this, we need to invoke the conditional distribution
of $\bB$ and $\bC$ with respect to the other ${\cal GP}$ parameters
and ${\cal D}_s$. We recall the 
priors on the ${\cal GP}$ parameters $\bB,\bSigma,\bC$ (from
Section~\ref{subsec:marginalized}) to write $\pi(\bB,\bSigma,
\bC)\propto\mid\bSigma\mid^{-(k+1)/2}\mid\bC\mid^{-(j+1)/2}$. It then
follows that
\begin{equation}
[\bB\mid\bSigma,\bC, \bQ,{\cal D}_s]\sim{\cal N}_{m,jk}(\hat
\bB_{GLS},(\bH_D^T\bA^{-1}_D\bH_D)^{-1},\bOmega), 
\label{eq:beta_posterior}
\end{equation}
where, we recall from Section~\ref{subsec:detailed} that we had set $m=d+1$, with $\bS\in{\mathbb R}^d$. Here, $\hat
\bB_{GLS}=(\bH_D^T\bA^{-1}_D\bH_D)^{-1}(\bH_D^T\bA^{-1}_D{\cal
  D}_s)$. Marginalising the $jk$-variate normal that is the
conditional $[\bxi(\cdot)\mid \bB, \bSigma,\bC,\bQ,{\cal D}_s]$ over
$\bB$ (using Equation~\ref{eq:beta_posterior}), it can be shown that
\begin{equation}
[\bxi(\cdot)\mid
\bSigma,\bC,\bQ,{\cal D}_s]\sim{\cal N}_{jk}(\bmu_2(\cdot),a_2(\cdot,\cdot)\bOmega),
\label{eq:posterior_predictive2}
\end{equation}
where
\begin{eqnarray}
\bmu_2(\cdot)&=&{{\hat \bB}_{GLS}}^T{\bh(\cdot)}+ ({\cal D}_s -\bH_D{\hat
\bB}_{GLS})^T\bA^{-1}_D\bsigma_D(\cdot) \label{eq:mean3};\\
a_2(\bs_1,\bs_2)&=&a_1(\bs_1,\bs_2)+[\bh(\bs_1)-\bH_D^T\bA^{-1}_D\bs_D(\bs_1)]^T
(\bH_D^T\bA^{-1}_D\bH_D)^{-1}\nonumber\\
&&[\bh(\bs_2)-\bH_D^T\bA^{-1}_D\bs_D(\bs_2)].
\label{eq:var3}
\end{eqnarray}
We define $(n-m){\hat{\bOmega}}_{GLS}= ({\cal D}_s -
\bH_D{\hat{\bB}}_{GLS})^T\bA^{-1}_D({\cal
  D}_s-\bH_D{\hat{\bB}}_{GLS})$,
i.e. $(n-m){\hat{\bOmega}}_{GLS}={\cal D}_s^T\bM{\cal D}_s$, with
$\bM=\bA^{-1}_D-\bA^{-1}_D\bH_D(\bH_D^T\bA^{-1}_D\bH_D)^{-1}\bH_D^T\bA^{-1}_D$).

We consider the mean $\bmu_2(\cdot)$ of the conditional posterior given by
(\ref{eq:mean3}) as a suitable estimator of the velocity function in
our case. Note that $\bmu_2$ involves the unknown smoothness
parameters; we plug-in the corresponding posterior
medians $0.874254,1.003545$ for these.

It is important to mention that though the mean and variance in
Equations~\ref{eq:mean3} and Equation~\ref{eq:var3} were developed
using ${\cal D}_s$, in our construction of the velocity function
estimator $\bmu_2$, ${\cal D}_a$ is implemented, where ${\cal D}_a$ is
obtained by augmenting ${\cal D}_s$ with $\bv^{(test)}$ that is
realised at $\bS=\tilde{\bs}$. The underlying theory remains the same
as above.

It is important to note that $\bmu_2(\bS)$, where $\bS$ is the unknown
location, is a random variable, and even though the posterior of
$\bSigma$ is concentrated around the null matrix, the variance of
$\bmu_2(\bS)$ is not $\bzero$, thanks to the fact that $\bS$ does not
have $\bzero$ variance. Consequently, the posterior variance of
$\bxi(\bS)$ does not have $\bzero$ variance. To see this
formally, note that
\begin{align}
Var\left[\bxi(\bS)\vert{\cal D}_a\right]
&= Var\left[{\mathbb E}\left\{\bxi(\bS)\vert\bSigma,\bC,\bQ,\bS,{\cal D}_a\right\}\right]
+ {\mathbb E}\left[Var\left\{\bxi(\bS)\vert\bSigma,\bC,\bQ,\bS,{\cal D}_a\right\}\right]\notag\\
&= Var\left[\bmu_2(\bS)\vert {\cal D}_a\right]+{\mathbb E}\left[a_2(\bS,\bS)\bOmega\vert {\cal D}_a\right].
\label{eq:condvar}
\end{align}
Since the posterior $[\bSigma\vert {\cal D}_a]$ is concentrated around
the $k\times k$-dimensional null matrix, it follows that the posterior
$[\bOmega\vert {\cal D}_a]$ is also concentrated around the $jk\times
jk$-dimensional null matrix. Hence, in (\ref{eq:condvar}),
${\mathbb E}\left[a_2(\bS,\bS)\bOmega\vert {\cal
    D}_a\right]\approx\bzero^{(jk\times jk)}$.  However, the first
part of (\ref{eq:condvar}), $Var\left[\bmu_2(\bS)\vert {\cal
    D}_a\right]$, is strictly (and significantly) positive, showing
that the variance of the posterior of $\bxi(\bS)$ is significantly positive.

The above result shows that it should not be expected that the
observed test velocity data $\bv^{(test)}$ will be predicted
accurately by $\bmu_2(\bs)$, for any given $\bs$.  This is in contrast
with the usual Gaussian process emulators, where the argument of the
unknown function is non-random, so that if the posterior of the
function variance is concentrated around $\bzero$, then the posterior
variance of the emulator would be close to $\bzero$.
 
\begin{figure}[!t]
    \begin{center}
{$\begin{array}{c c c c}
\includegraphics[width=3.5cm,height=4cm]{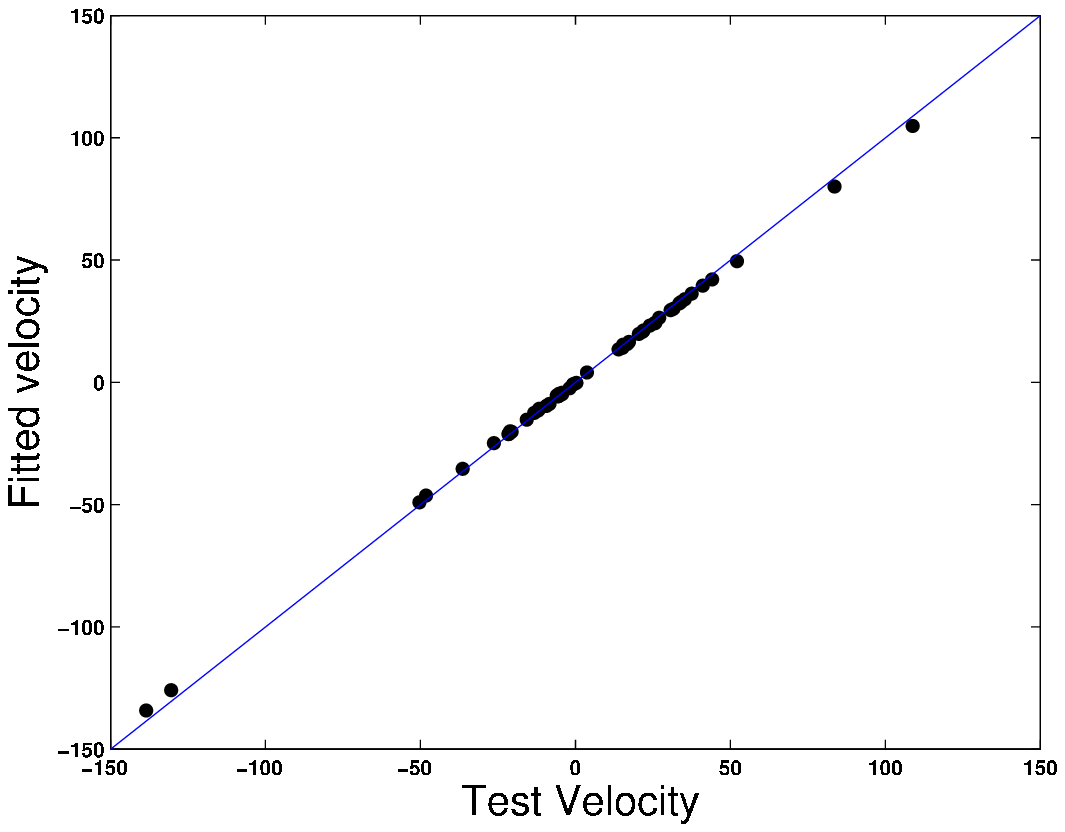} 
\includegraphics[width=3.5cm,height=4cm]{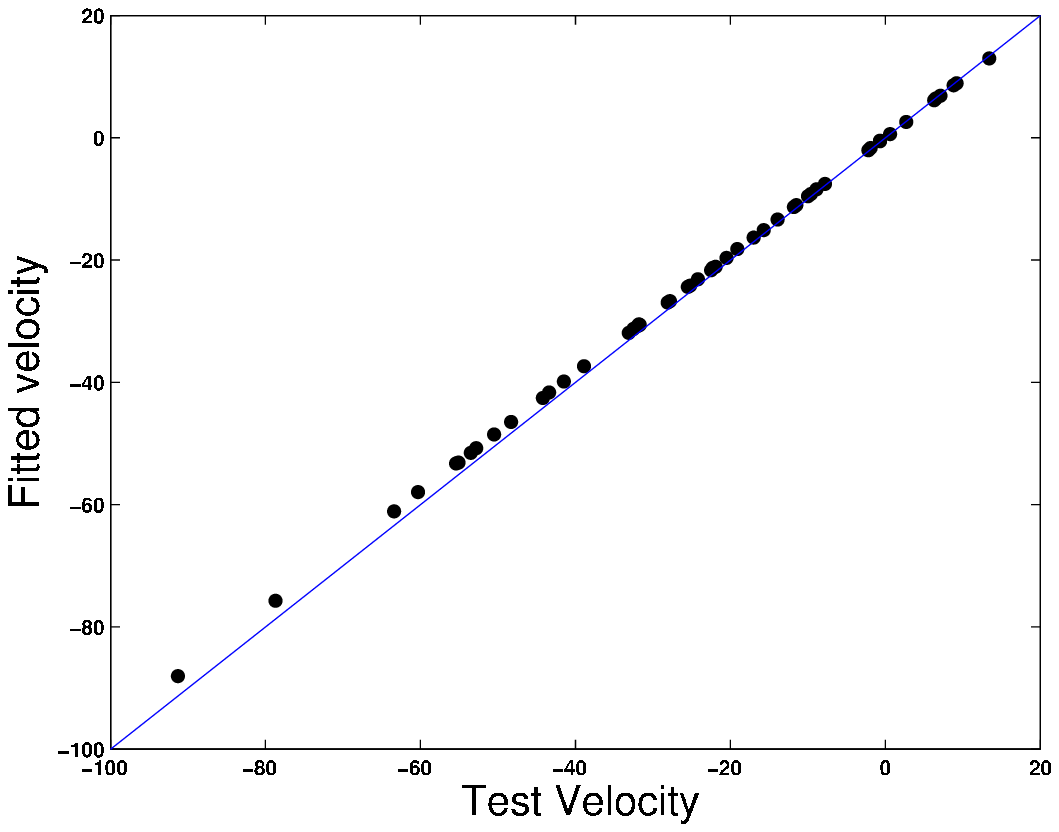}\quad\quad 
\includegraphics[width=3.5cm,height=4cm]{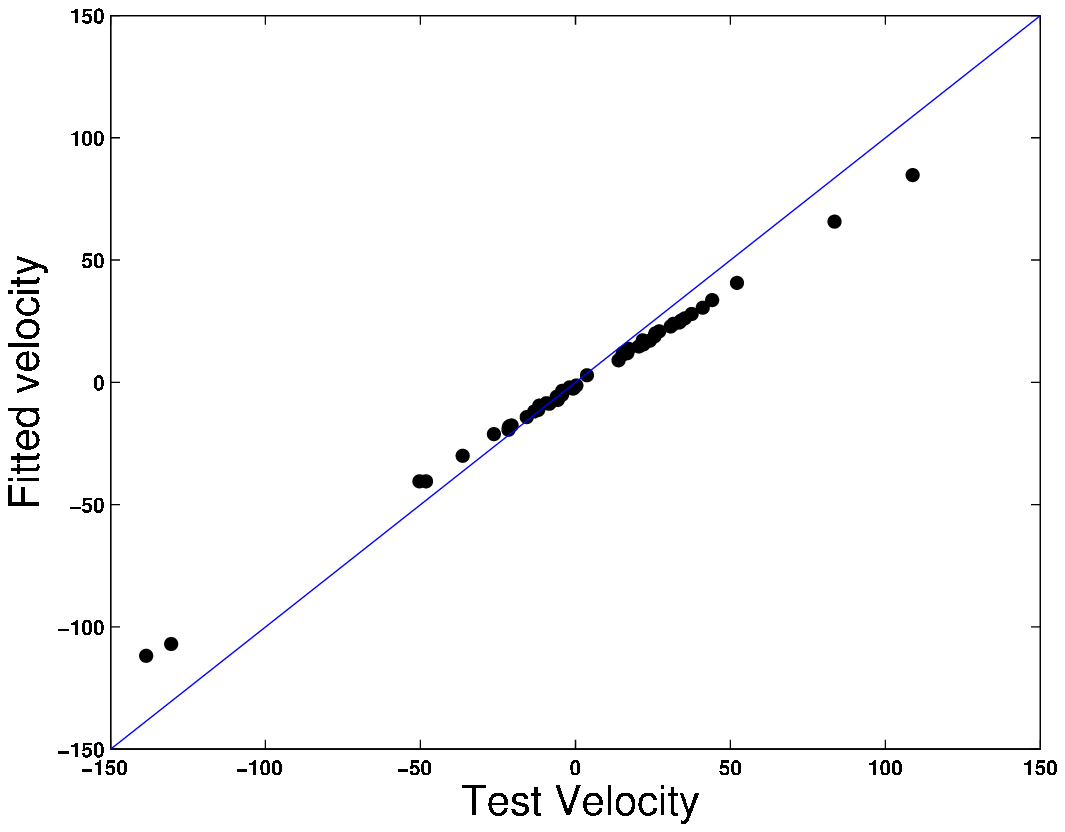} 
\includegraphics[width=3.5cm,height=4cm]{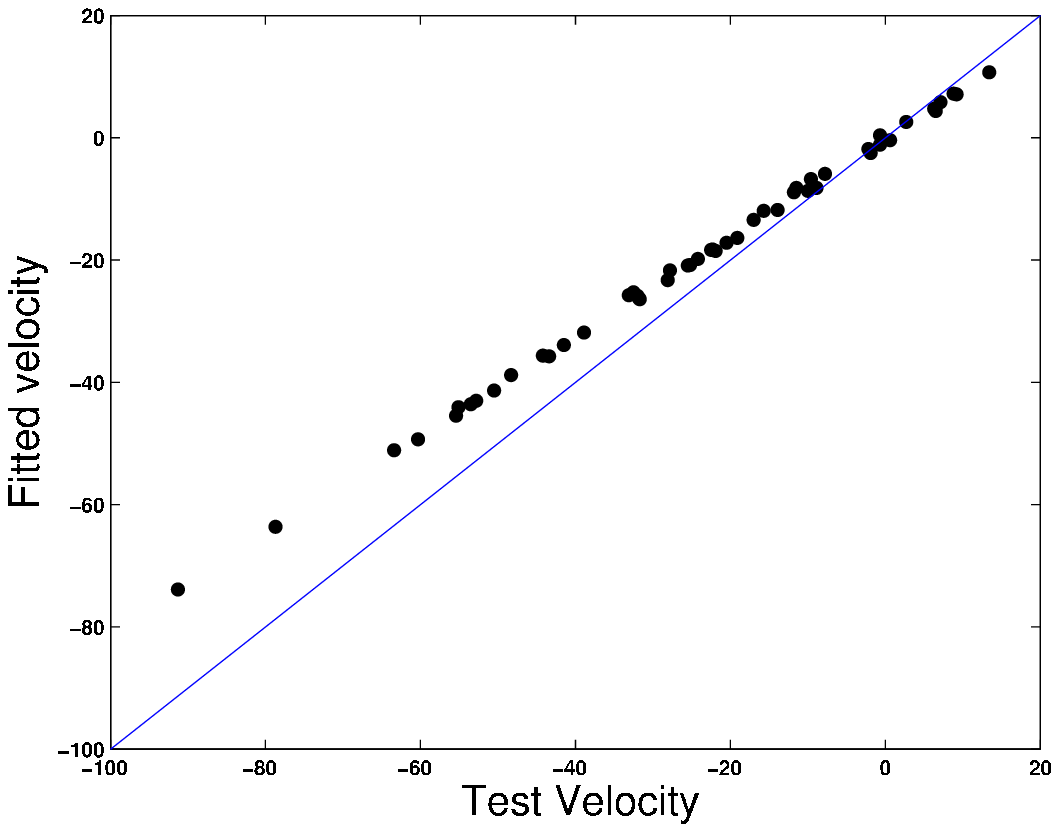}\\
\includegraphics[width=3.5cm,height=4cm]{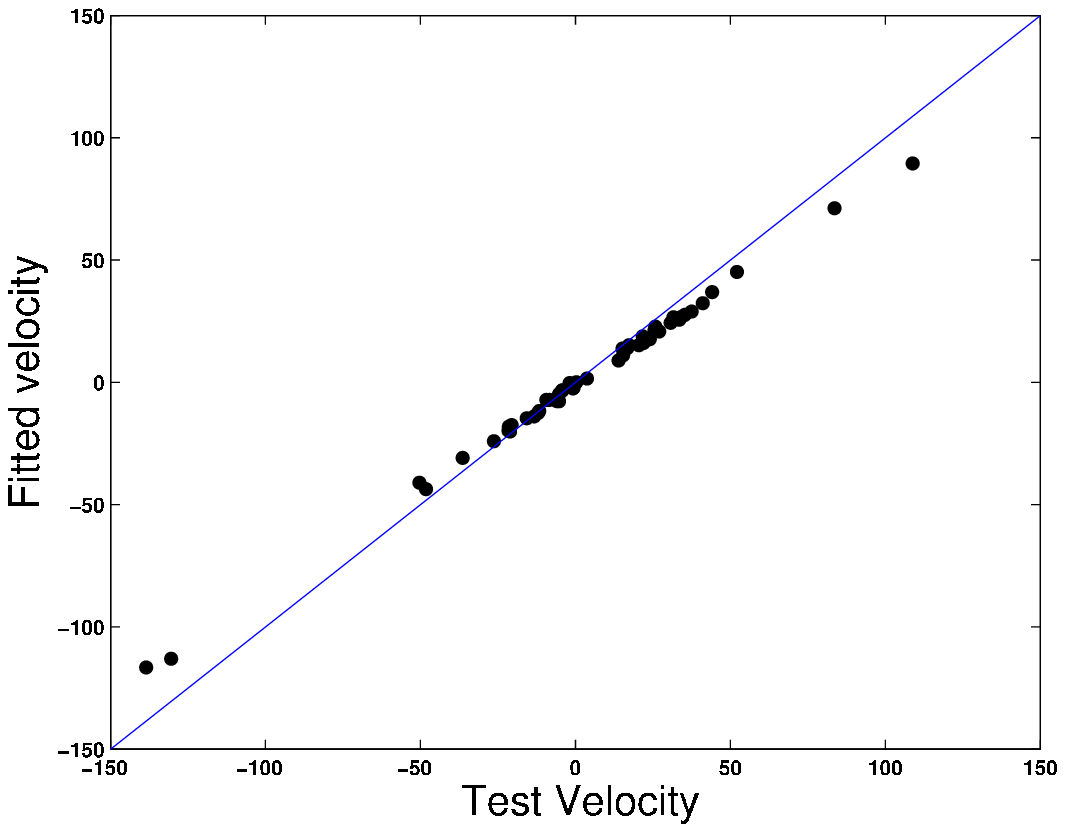}
\includegraphics[width=3.5cm,height=4cm]{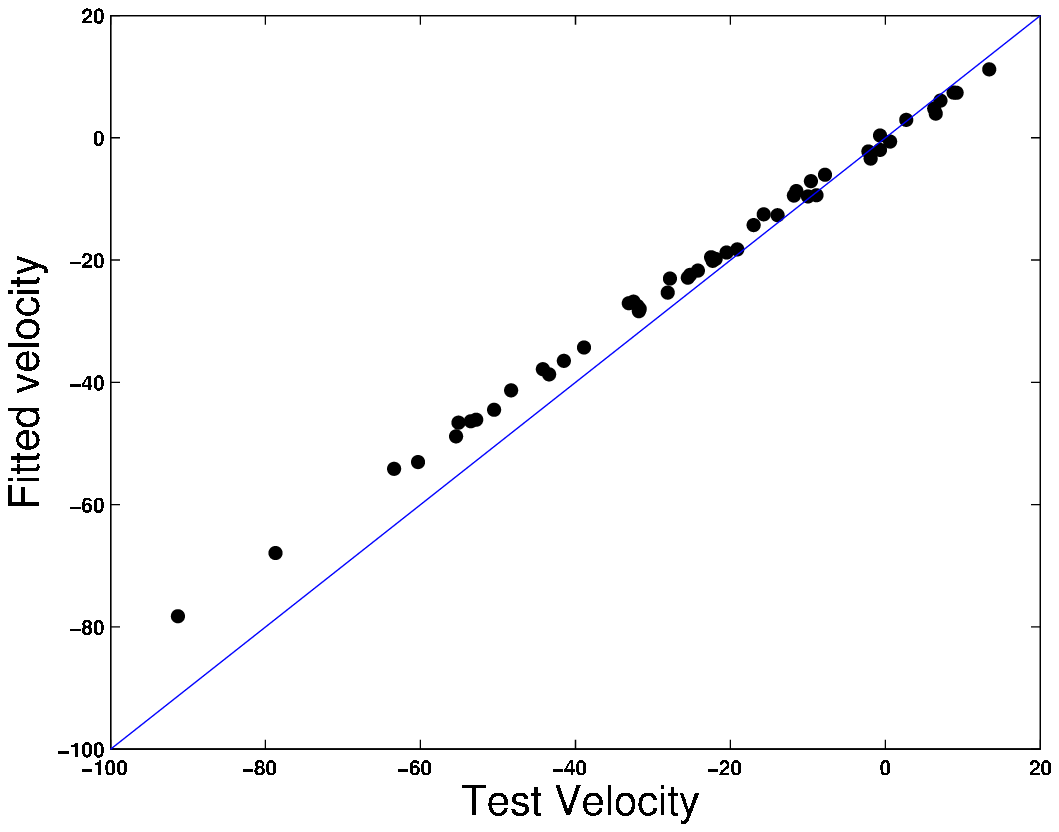}\quad\quad
\includegraphics[width=3.5cm,height=4cm]{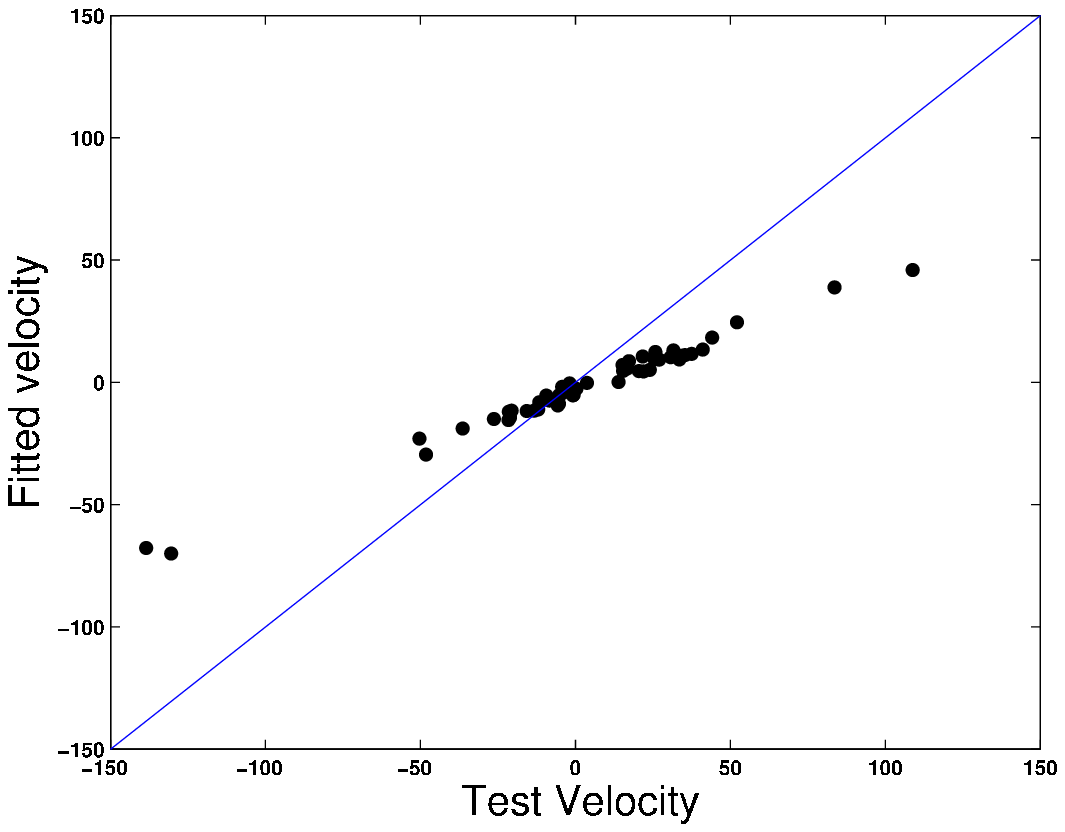} 
\includegraphics[width=3.5cm,height=4cm]{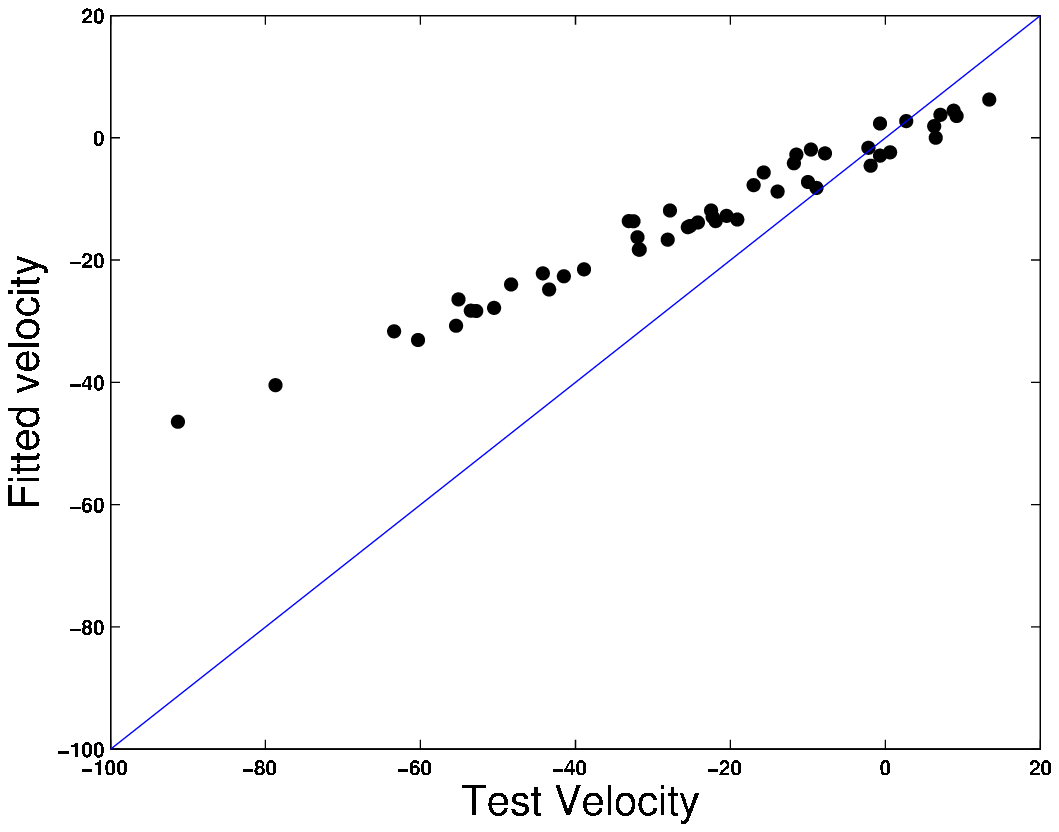} \\
\includegraphics[width=3.5cm,height=4cm]{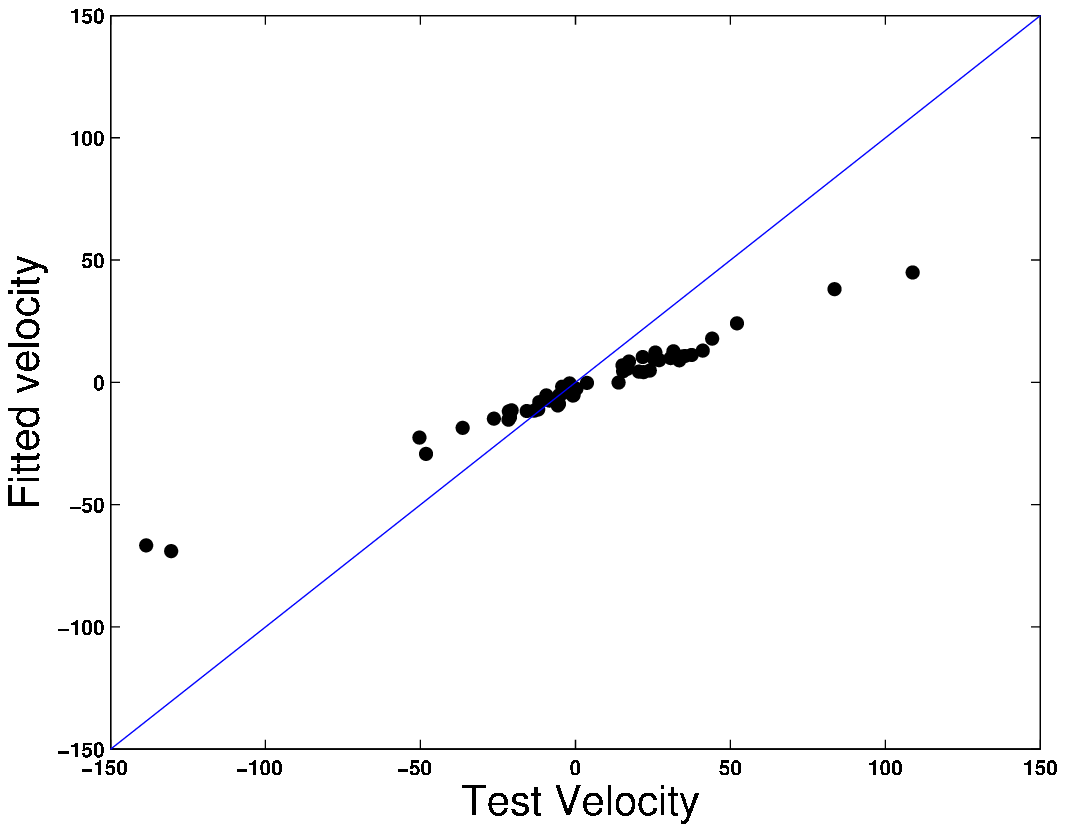} 
\includegraphics[width=3.5cm,height=4cm]{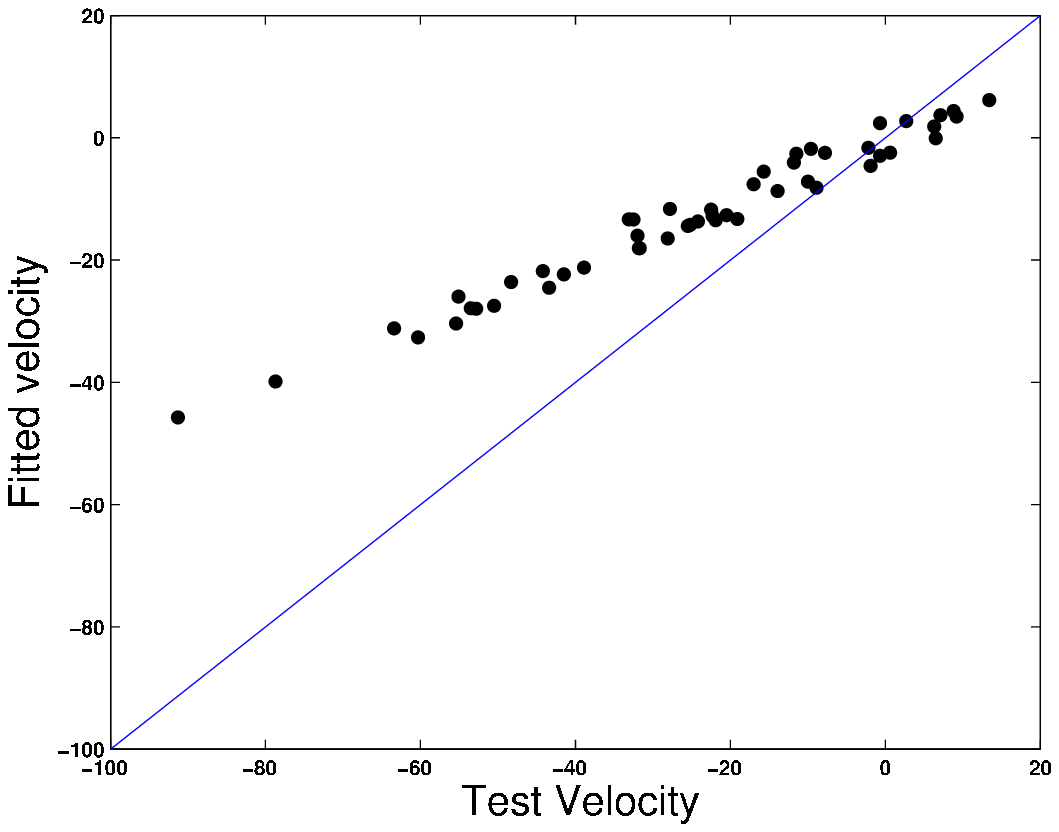}\quad\quad
\includegraphics[width=3.5cm,height=4cm]{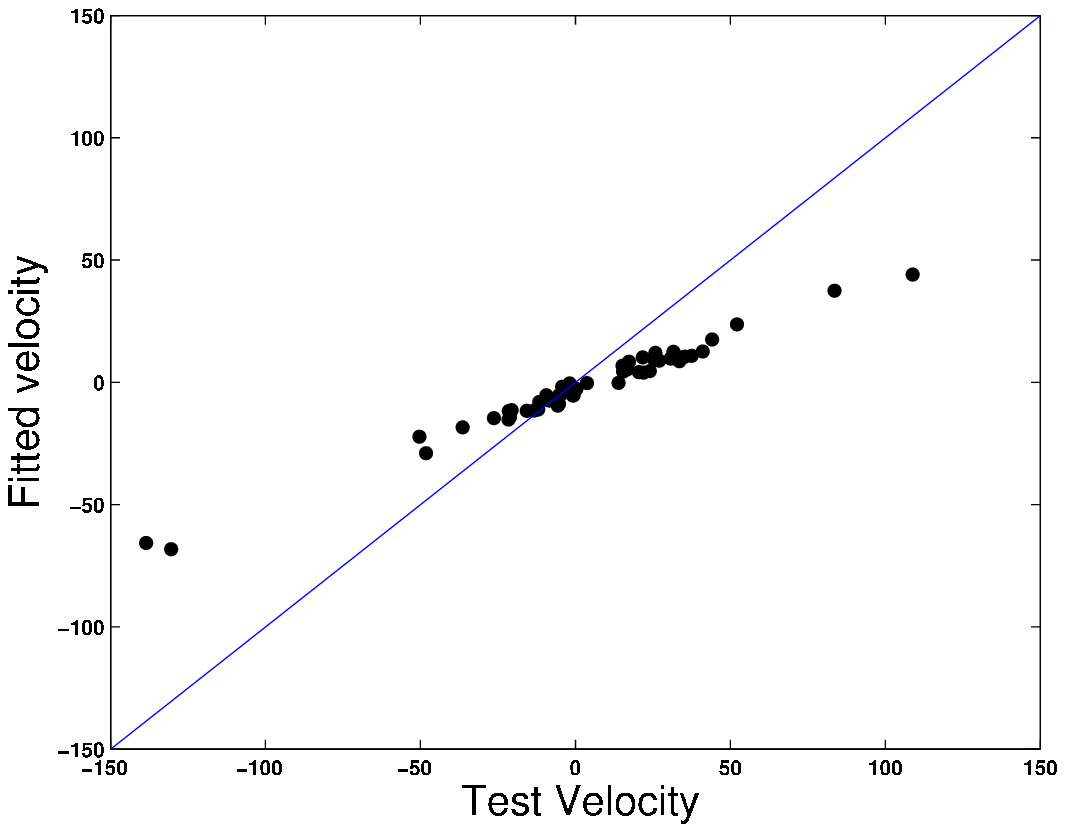}
\includegraphics[width=3.5cm,height=4cm]{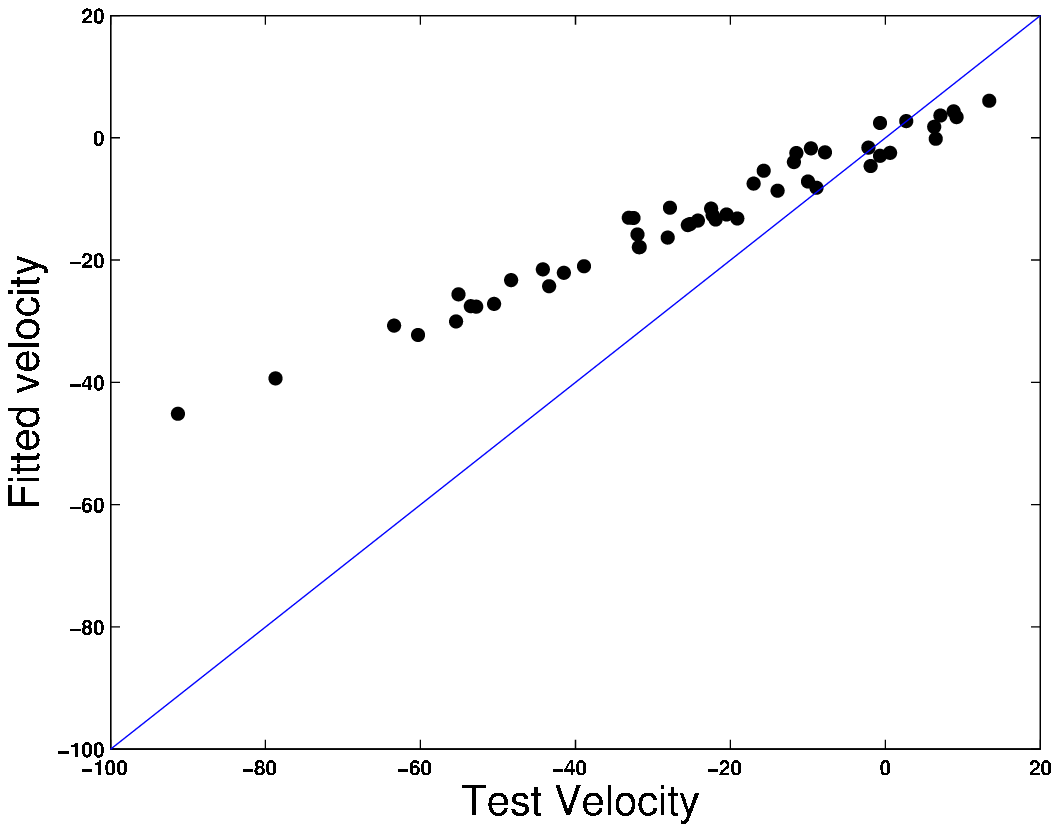} 
\end{array}$}
\end{center}
\caption{Prediction of $\bv^{(test)}$ for model $sp3bar3$: plots of 2 components of $\bmu_2(\bs)$ against $\bv^{(test)}$ for $\bs={\tilde{\bs}}$ (2 left hand sided panels on the top row), $bs=\bs^{(mode)}$ (2 right panels on the top), $\bs=\bs^{(1)}$ (2 left panels in the middle row), $\bs=\bs^{(2)}$ (2 left panels in the middle row), $\bs=\bs^{(3)}$ (2 left panels in the lowest row), $\bs=\bs^{(4)}$ (2 right panels in the lowest row).}
\label{fig:modelfit_1}
\end{figure}

In Figure~\ref{fig:modelfit_1} we
illustrate, in the case of $sp3bar3$ (the most chaotic model), the
degree of agreement of $\bmu_2(\bs)$ with $\bv^{(test)}$ for different
choices of $\bs$. 
We compare with $\bv^{(test)}$ the predictions
$\bmu_2(\bs^{(mode)})$, $\bmu_2({\tilde{\bs}})$ and $\bmu_2(\bs^{(u)});u=1,2,3,4$, 
Here ,
$\bs^{(mode)}=(1.73,18.8^\circ)$ is the (component-wise) posterior
mode and ${\tilde{\bs}}=(2.2,35^\circ)$ is a point somewhat close to the
(component-wise) posterior median
$\bs^{(median)}=(1.994478,33.59429^\circ)$ (grid-point closest to $\bs^{(median)}$.

As observed in Figure \ref{fig:modelfit_1} 
the best fit of $\bv^{(test)}$ has been provided by $\bmu_2({\tilde{\bs}})$
where ${\tilde{\bs}}$ is close to the median $\bs^{(median)}$; as the
point $(\bs^{(median)},\bv^{(test)})$ is in the training data
constituting $\bmu_2$, this is to be expected. The estimators
$\bmu_2(\bs^{(mode)})$ and $\bmu_2(\bs^{(1)})$ perform somewhat
reasonably, but the remaining estimators $\bmu_2(\bs^{(u)});u=2,3,4$
do not perform adequately, signifying the effect of variablity of our
estimator due the posterior of $\bS$.

\begin{figure}[!bh]
    \begin{center}
{$\begin{array}{c c c c}
\includegraphics[width=3.5cm,height=4cm]{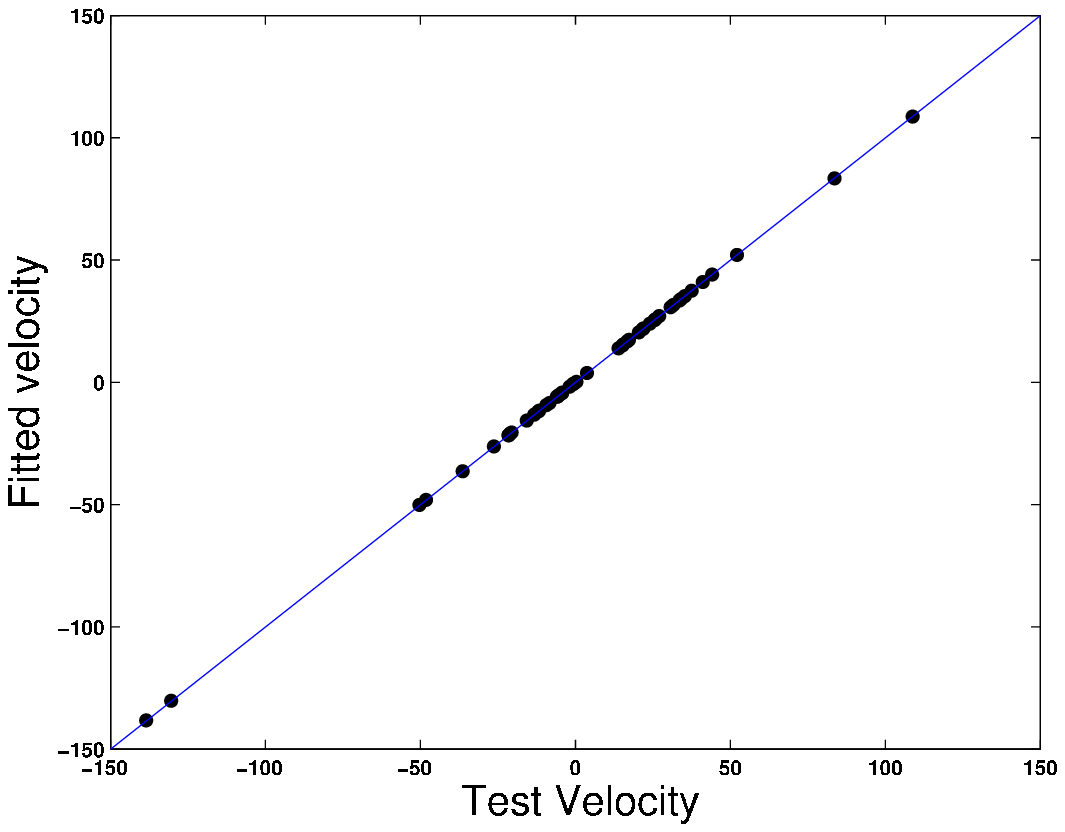}  
\includegraphics[width=3.5cm,height=4cm]{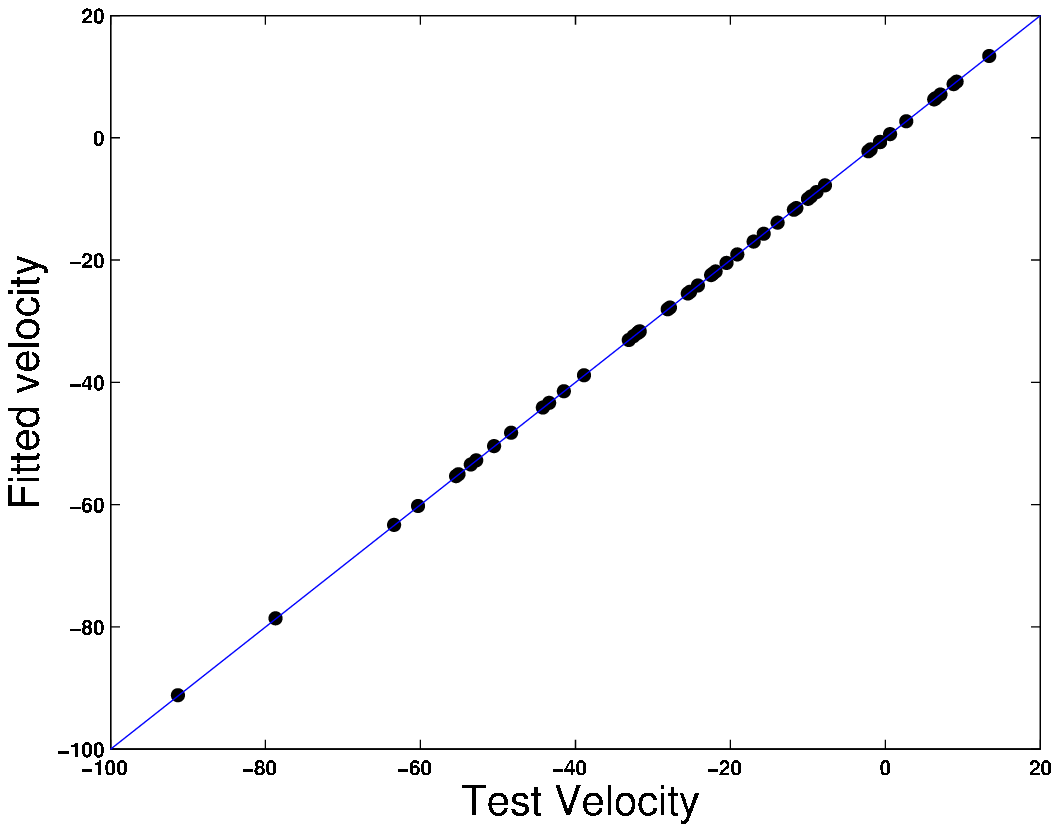} \quad\quad

\includegraphics[width=3.5cm,height=4cm]{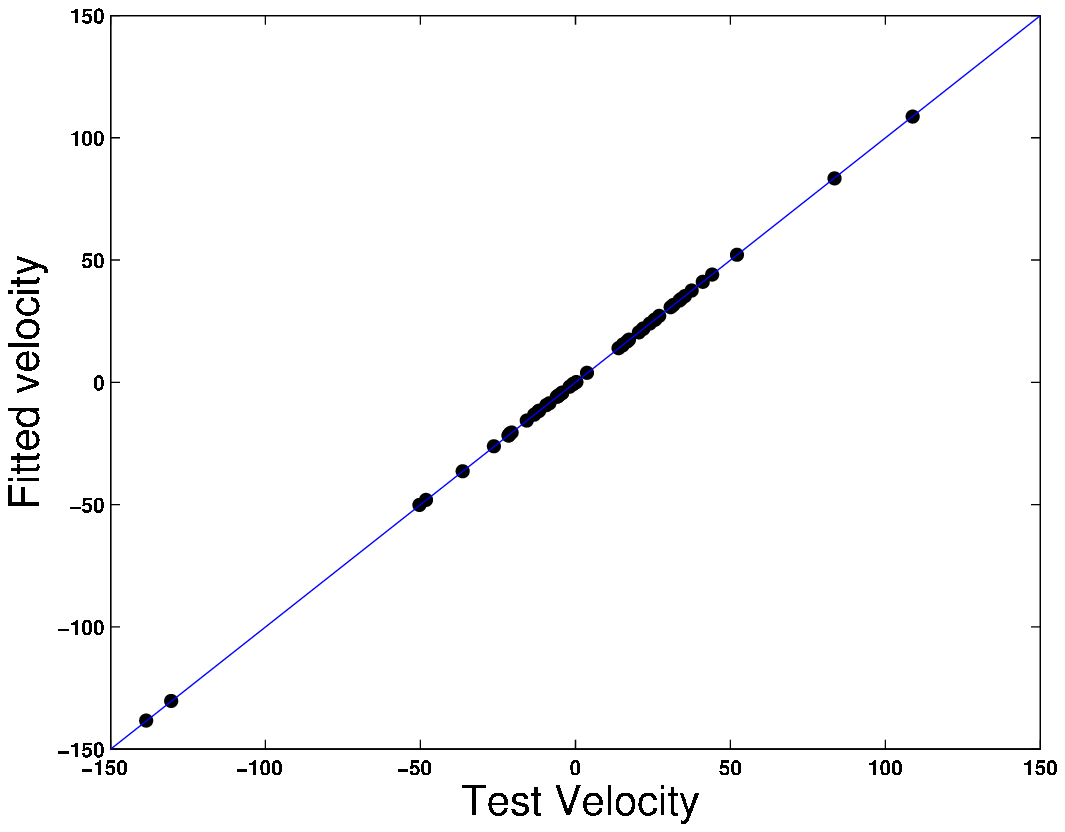} 
\includegraphics[width=3.5cm,height=4cm]{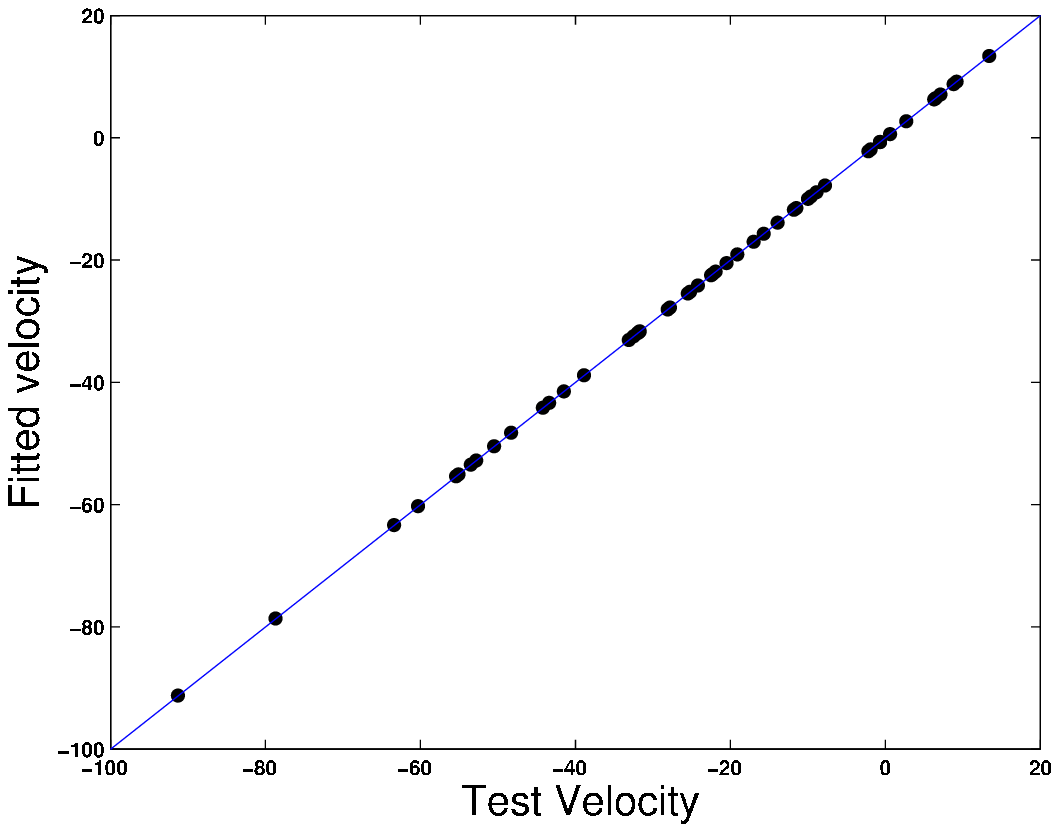} \\

\includegraphics[width=3.5cm,height=4cm]{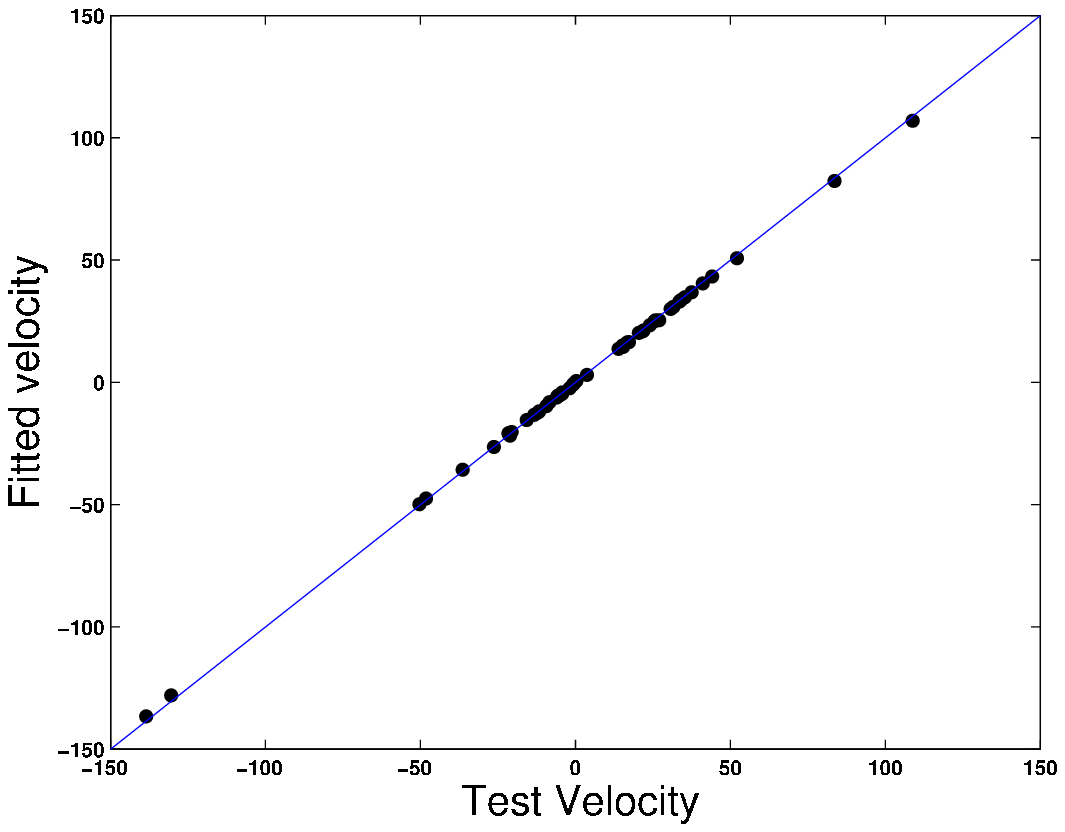} 
\includegraphics[width=3.5cm,height=4cm]{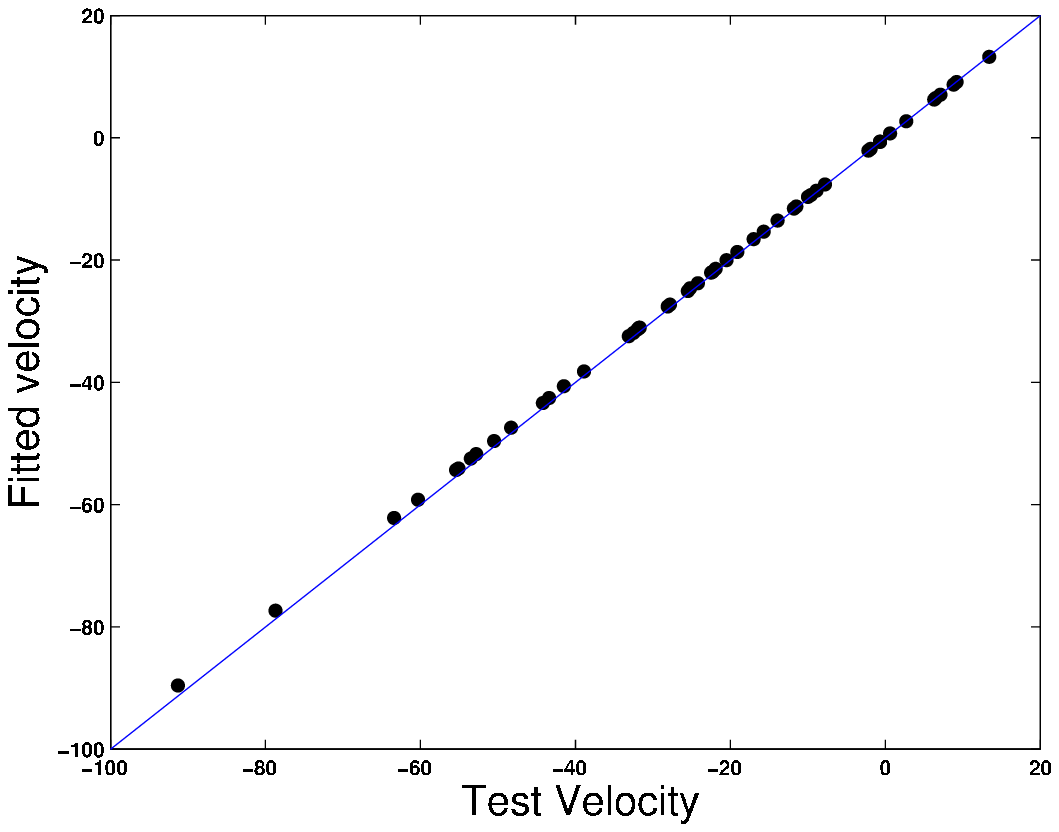} \quad\quad

\includegraphics[width=3.5cm,height=4cm]{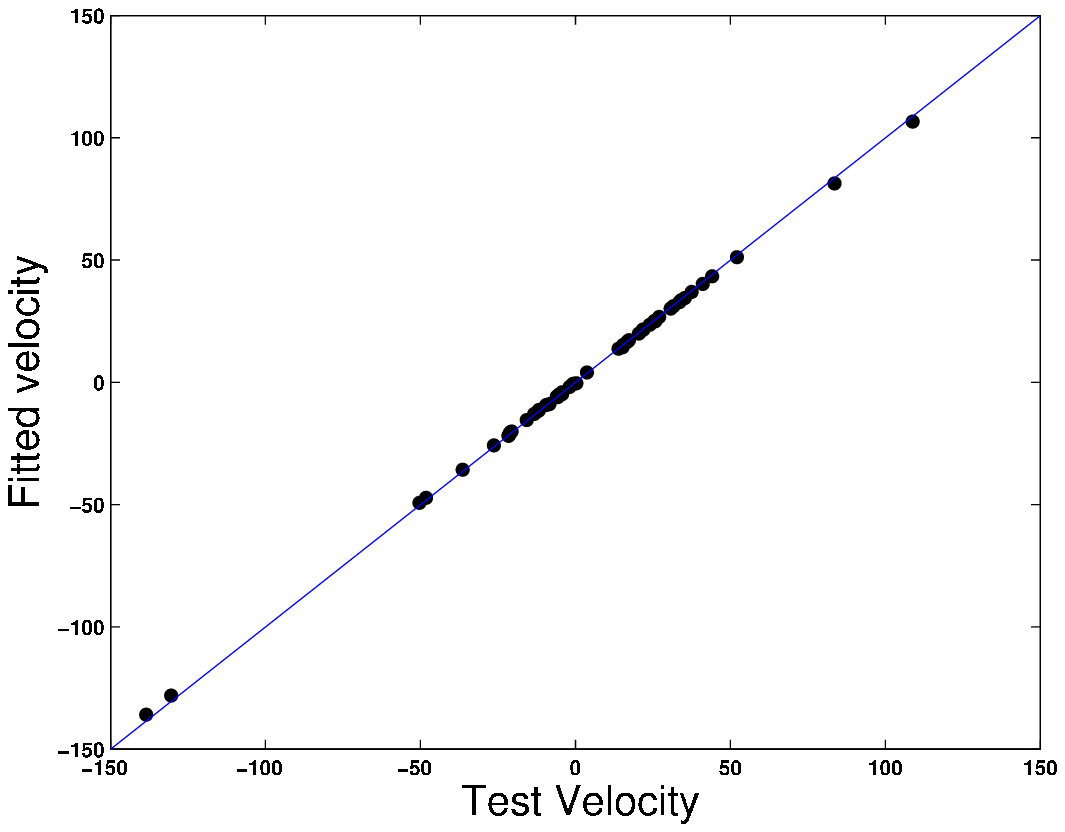} 
\includegraphics[width=3.5cm,height=4cm]{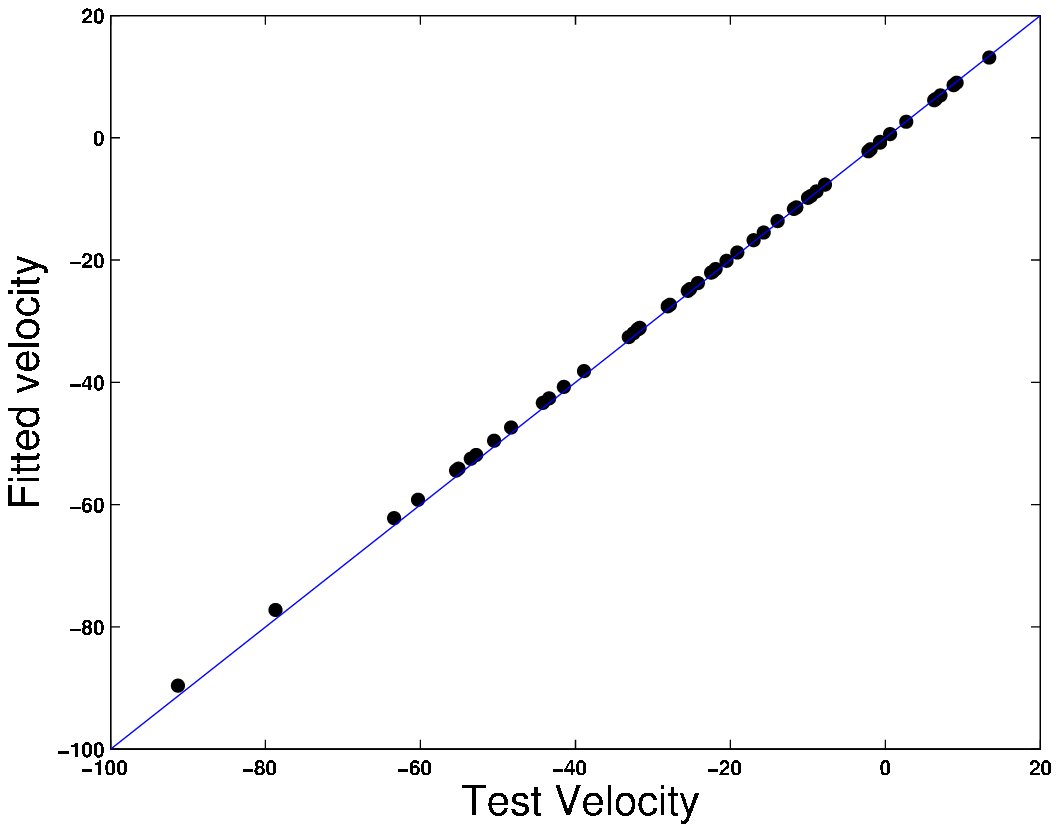}
\end{array}$
}
\end{center}
\caption{Prediction of $\bv^{(test)}$ for model $bar6$: plots of 2 components of $\bmu_2(\bs)$ against $\bv^{(test)}$ for $\bs={\tilde{\bs}}$ (2 adjacent panels on the left hand side of the top row), $\bs^{(mode)}$ (2 panels on the right of top row), $\bs^{(1)}$ (2 panels on the left in the lower row), $\bs^{(2)}$ (2 panels on the left in the lower row).}
\label{fig:modelfit_bar1}
\end{figure}

While it is the randomness of the argument $\bS$ of the unknown
function $\bxi(\cdot)$ that causes the variability of our estimator,
such variability is highest in the most chaotic of the 4 base
astrophysical models ($sp3bar3$), and least in the only non-chaotic
base astrophysical model ($bar\_6$). A similar exercise of predicting
$v^{(test)}$ using the training data simulated from this non-chaotic
base model gives excellent fits at all the aforementioned used values
of $\bS$; see Figure~\ref{fig:modelfit_bar1}.

\section{Discussions}
\label{sec:discussions}
\noindent

Computational complexity
scales only linearly with the dimensionality of the unknown model
parameter $\bS$. Thus, porting a training data comprised of $n$
independent values of $\bS$, $\bs_i,\:i=1,\ldots,n$, where $\bs_i$ is
a $d$-dimensional vector, $d>2$, is not going to render the
computational times infeasible. This allows for the learning of
high-dimensional model parameter vectors in our method.

In contrast to the situation with increasing the dimensionality of the
unknown model parameter, increasing the dimensionality of the
measurable will but imply substantial increase in the run time, since
the relevant computational complexity then scales non-linearly, as
about $O(k^3)$, (in addition to the cost of $k$ square roots), where
$k$ is the dimensionality of the observed variable. This is because of
the dimensionality of the aforementioned ${\bSigma}$ matrix is
$k\times k$, and the inverse of this enters the computation of the
posterior via the definition ${\hat{\bC}}_{GLS,aug}$. Thus, for
example, increasing the dimensions of the measurable from 2 to
4 increases the run time 8-fold, which is a large
jump in the required run time. However, for most applications, we
envisage the expansion of the dimensionality of the unknown model
parameter, i.e. $d$, rather than that of the measurable,
i.e. $k$. Thus, the method is expected to yield results within
acceptable time frames, for most practical applications.

The other major benefit of our work is that it allows for organic
learning of the smoothness parameters, rather than results being
subject to {\it ad hoc} choices of the same.

As more Galactic simulations spanning a greater range of model
parameters become available, the rigorous learning of such Milky Way
parameters using our method will become possible, given the available
stellar velocity data. This will enhance the quality of our knowledge
about our own galaxy. That our method allows for such learning even
for under-abundant systems, is encouraging for application of a similar
analysis to galaxies other than our own, in which system parameters
may be learnt using the much smaller available velocity data sets,
compared to the situation in our galaxy. 

\begin{center}
{\bf{Supplementary material}}
\end{center}
{ Some background details on the application to the Milky Way are
  discussed in Section {\bf{S-1}} of the attached supplementary
  material.  Section {\bf{S-2}} discusses the details of the dynamical
  simulations that lead to the training data set used in our
  supervised learning of the Milky Way feature parameters.  In
  Section~{\bf{S-3}} we present details of the TMCMC methodology that
  we use here. {\bf{S-4}}
  discusses the cross-validation of our model and methodology, on
  simulated as well the real stellar velocity data. The effect of
  chaos on the modality of the posterior distributions of our unknowns
  is discussed in Section~{\bf{S-5}}.}


\renewcommand\baselinestretch{1.}
\small
\bibliographystyle{ECA_jasa}
\bibliography{irmcmc_sou}
\end{document}